\documentclass[11pt]{article}

\PassOptionsToPackage{subfigure}{tocloft}
\usepackage{graphicx}
\usepackage{subfigure}
\usepackage[space]{grffile}
\usepackage{simplewick}
\usepackage{array}
\usepackage{color}
\usepackage{dsdshorthand}
\usepackage{mathtools}
\usepackage{float}
\usepackage{tabularx}  
\usepackage{chet}
\usepackage{dsfont}
\usepackage{amsmath}

\definecolor{darkblue}{rgb}{0.1,0.1,0.7}
\hypersetup{colorlinks,
           linkcolor={darkblue},
           citecolor={darkblue},
           urlcolor={darkblue}
}

\newcommand{\overbar}[1]{\mkern 3.5mu\overline{\mkern-3.5mu#1\mkern-1.5mu}\mkern 1.5mu}
\newcommand{\overbarUp}[1]{\mkern 1.5mu\overline{\mkern-1.5mu#1\mkern-2mu}\mkern 2mu}
\newcommand{\lifrac}[2]{\mbox{$\frac{#1}{#2}$}}

\newcommand{\ib}{{\bar{\imath}}}
\newcommand{\bfz}{\mathbf{z}}
\newcommand{\chib}{\bar{\chi}}
\newcommand{\Zb}{\overbar{Z}}
\newcommand{\cOb}{{\overbar{\cO}}}
\newcommand{\rmx}{{\mathrm{x}}}
\newcommand{\tr}{{\mathrm{tr}\,}}
\newcommand{\Thetab}{{\overbarUp\Theta}}
\newcommand{\CN}{\mathcal{N}}
\newcommand{\CO}{\mathcal{O}}
\newcommand{\COb}{{\overbar{\mathcal{O}}}}
\newcommand{\Ob}{\overbar{O}}
\newcommand{\CJ}{\mathcal{J}}

\newcommand{\Qb}{{\overbar{Q}}{}}

\newcommand{\qb}{{\bar{q}}}

\newcommand{\xb}{{\bar{x}}}
\newcommand{\di}{\mathrm{d}}
\newcommand{\jb}{{\bar{\jmath}}\hspace{0.9pt}}

\newcommand{\etab}{{\bar{\eta}}}
\newcommand{\thetab}{{\bar{\theta}}}

\newcommand{\sigmab}{{\bar{\sigma}}}
\newcommand{\Xb}{{\overbar{X}}{}}
\newcommand{\alphad}{{\dot{\alpha}}}

\newcommand{\betad}{{\smash{\dot{\beta}}}}

\newcommand{\vev}[1]{{\langle #1\rangle}}
\newcommand{\Vev}[1]{\left\langle #1\right\rangle}

\newcommand{\lsp}{\hspace{1pt}}
\newcommand{\llsp}{\hspace{0.5pt}}
\newcommand{\lnsp}{\hspace{-0.8pt}}

\newcommand{\sectionname}{Sec.} 

\newcommand{\beq}{\begin{equation}}
\newcommand{\eeq}{\end{equation}}
\newcommand{\eref}[1]{(\ref{#1})}

\newlength{\mytextheight}
\newsavebox{\mytext}
\newcommand{\tableeqn}[1]{\savebox{\mytext}{$\displaystyle #1$}
                          \settoheight{\mytextheight}{\usebox{\mytext}}
                          {\raisebox{\the\mytextheight+.5ex}{}\hspace{-1.8ex}\vspace{.25ex}$\displaystyle #1$}
                          }

\newcounter{appendixfloats}
\newcommand{\appendset}[2]
{\stepcounter{appendixfloats}%
  \expandafter\gdef\csname #2\theappendixfloats\endcsname{#1}%
}
\newcommand{\appendlist}[1]
{\bgroup
  \count1=0
  \loop\ifnum\count1<\value{appendixfloats}\relax
    \advance\count1 by 1
    \csname #1\the\count1\endcsname
  \repeat
\egroup}

\newcommand{\cOJOA}{{\cC_1}}
\newcommand{\cOJOB}{{\cC_2}}
\newcommand{\cOJOC}{{\cC_3}}
\newcommand{\cOJOD}{{\cC_4}}
\newcommand{\cOJOE}{{\cC_5}}
\newcommand{\cOJOF}{{\cC_6}}
\newcommand{\cOJOG}{{\cC_7}}
\newcommand{\cOJOH}{{\cC_8}}
\newcommand{\cOJOI}{{\cC_9}}
\newcommand{\cOJOJ}{{\cC_{10}}}
\newcommand{\cOJOi}[1]{{\cC_{#1}}} 
\newcommand{\cObJO}[1]{{C_{#1}}}
\newcommand{\cObTO}[1]{{D_{#1}}}
\newcommand{\cQObTQbO}[1]{{H_{#1}}}
\newcommand{\cQObSbO}[1]{{G_{#1}}}
\newcommand{\cQbObSOplus}[1]{{E_{#1}}}
\newcommand{\cQbObSOminus}[1]{{F_{#1}}}
\newcommand{\cQbObJQOmm}[1]{{Q_{#1}}}
\newcommand{\cQbObJQOmp}[1]{{P_{#1}}}
\newcommand{\cQbObJQOpm}[1]{{O_{#1}}}
\newcommand{\cQbObJQOpp}[1]{{N_{#1}}}
\newcommand{\cQObJQbO}[1]{{I_{#1}}}
\newcommand{\cQbObTQOmm}[1]{{M_{#1}}}
\newcommand{\cQbObTQOmp}[1]{{L_{#1}}}
\newcommand{\cQbObTQOpm}[1]{{K_{#1}}}
\newcommand{\cQbObTQOpp}[1]{{J_{#1}}}

\def\ge{\geqslant}

\def\geq{\geqslant}
\def\leq{\leqslant}

\newcommand{\Uc}[2]{\lsp [#1\ifthenelse{\equal{#2}{\Theta}}{\Thetab}{\bar{#2}}]}
\newcommand{\Ec}[2]{\lsp [#1#2]}
\newcommand{\Ecb}[2]{\lsp [\ifthenelse{\equal{#1}{\Theta}}{\Thetab}{\bar{#1}}
                           \ifthenelse{\equal{#2}{\Theta}}{\Thetab}{\bar{#2}}]}

\newcommand{\invI}[2]{\mathbb{I}^{\lsp#1#2}}
\newcommand{\invJ}[3]{\mathbb{J}^{\lsp#3}_{#1#2}}
\newcommand{\invK}[3]{\mathbb{K}_{\lsp#3}^{#1#2}}
\newcommand{\invKb}[3]{\overbarUp{\mathbb{K}}_{\lsp#3}^{#1#2}}

\newcommand{\susyT}[1]{\mathbb{T}_{#1}}
\newcommand{\nonsusyT}[1]{\cT_{#1}}

\newcommand{\anecE}[3]{\cE[#1;(#2);#3]}
\newcommand{\anecEint}[3]{\cE_\mathrm{int}[#1;(#2);#3]}

\title{Implications of ANEC for SCFTs in four dimensions}

\author{Andrea Manenti,${\!}^{a,b}$ Andreas Stergiou,${\!}^c$
and Alessandro Vichi${}^a$}

\affiliation{${}^a$Institute of Physics, \'Ecole Polytechnique F\'ed\'erale de Lausanne, CH-1015 Lausanne, Switzerland\\\vspace{2pt}
${}^b$Simons Center for Geometry and Physics, Stony Brook, New York,
USA\\\vspace{2pt}
${}^c$Theoretical Division, MS B285, Los Alamos National Laboratory, Los Alamos, NM 87545, USA}

\abstract{We explore consequences of the Averaged Null Energy Condition
(ANEC) for scaling dimensions $\Delta$ of operators in four-dimensional
$\mathcal{N}=1$ superconformal field theories. We show that in many cases
the ANEC bounds are stronger than the corresponding unitarity bounds on
$\Delta$. We analyze in detail chiral operators in the $(\frac12 j,0)$
Lorentz representation and prove that the ANEC implies the lower bound
$\Delta\ge\frac32j$, which is stronger than the corresponding unitarity
bound for $j>1$. We also derive ANEC bounds on $(\frac12 j,0)$ operators
obeying other possible shortening conditions, as well as general $(\frac12
j,0)$ operators not obeying any shortening condition. In both cases we find
that they are typically stronger than the corresponding unitarity bounds.
Finally, we elucidate operator-dimension constraints that follow from our
$\mathcal{N}=1$ results for multiplets of $\mathcal{N}=2,4$ superconformal
theories in four dimensions. By recasting the ANEC as a convex optimization
problem and using standard semidefinite programming methods we are able to
improve on previous analyses in the literature pertaining to the
nonsupersymmetric case.}

\date{May 2019}

\begin{document}

\maketitle

\toc

\section{Introduction and summary of results}
In recent years attention has been brought to the utility of expectation
values of integrated projections of the stress-energy tensor along null
lines in conformal field theories (CFTs). Such observables have a long
history in jet physics---see for example~\cite{Sveshnikov:1995vi,
Belitsky:2001ij, Lee:2006fn}---and they were first examined in the CFT
context in the seminal work \cite{Hofman:2008ar}. There, it was shown that
an energy-positivity condition implies constraints on the coefficients in
the three-point function of the stress-energy tensor. More precisely, given
a state $|\psi\rangle$ of a local CFT with stress-energy tensor
$T_{\mu\nu}$ and a null geodesic parametrized by $\lambda$ with tangent
vector $u^\mu$, the following inequality, called the Averaged Null Energy
Condition (ANEC), holds:
\eqn{
 \langle\psi |\mathcal E |\psi\rangle =  \int_{-\infty}^\infty \di\lambda\, \langle\psi |T_{\mu\nu}|
 \psi\rangle\,u^\mu u^\nu  \geq 0\,.
}[eq:ANEC]
In~\cite{Hofman:2008ar} this was viewed as a positivity requirement for the
energy measured by a hypothetical ``calorimeter'' placed at a large
distance from the region where $|\psi\rangle$ is localized. The
inequality~\eqref{eq:ANEC} was later established more rigorously as a
theorem~\cite{Faulkner:2016mzt,Hartman:2016lgu}. It has also been
understood that the ANEC is part of a larger class of constraints, which
also follow from a thought collider experiment, namely the deep inelastic
scattering bounds~\cite{Komargodski:2016gci, Meltzer:2018tnm}, which state
the positivity of an expectation value similar to \eqref{eq:ANEC} but with
$T_{\mu\nu}$ replaced by the lowest-twist operator of a given spin $\ell >
2$. Recently it was shown that the integral \eref{eq:ANEC} is a special
case of a larger set of integral transforms~\cite{Kravchuk:2018htv}.

An important, perhaps unexpected application of \eqref{eq:ANEC} is that it
implies lower bounds on operator dimensions $\Delta$ in
CFTs~\cite{Cordova:2017dhq}. It is of course known that in CFTs scaling
dimensions of operators are bounded from below as a consequence of
unitarity~\cite{Mack:1975je, Grinstein:2008qk}.  This is true independently
of locality properties of the CFT, i.e.\ it does not rely on the presence
of a stress-energy tensor in the CFT spectrum.  However, it was
demonstrated in \cite{Cordova:2017dhq} that in CFTs with a stress-energy
tensor the unitarity bound is suboptimal for some classes of operators. The
analysis of a few examples led~\cite{Cordova:2017dhq} to the conjecture
$\Delta\ge\max\{j,\jb\}$, where $(\frac12 j,\frac12\jb)$ is the Lorentz
representation of the CFT operator. This becomes stronger than the
unitarity bound whenever $|j-\jb|>4$. We find that this conjecture is not
supported by the ANEC for large values of $j$ in the case of $(\frac12
j,0)$ and $(\frac12 j,\frac12)$ operators---see Figs.~\ref{fig:nonsusyJ0}
and~\ref{fig:nonsusyJ1} below.

In this work we apply the methods of~\cite{Cordova:2017dhq} to
four-dimensional $\cN=1$ superconformal field theories (SCFTs). We find
that for certain classes of operators the unitarity bounds of
\cite{Flato:1983te, Dobrev:1985qv, Minwalla:1997ka} cannot be saturated.
Just as in~\cite{Cordova:2017dhq}, our results follow from a careful
analysis of three-point functions of the schematic type $\vev{\Ob\llsp
T_{\mu\nu}O}$ with $O$ a conformal primary and $\Ob$ its conjugate. The
difference with the nonsupersymmetric case is that here such conformal
three-point functions are encoded in superconformal three-point functions
involving the Ferrara--Zumino multiplet~\cite{Ferrara:1974pz}. The
constraints of 4d $\CN=1$ superconformal symmetry on three-point functions
of superconformal primary operators have been examined in great detail
in~\cite{Park:1997bq, Osborn:1998qu}, and we rely heavily on those results.

The constraints imposed by the ANEC and explored in \cite{Cordova:2017dhq} are
schematically of the form
\eqn{
\Delta_{O} > \Delta_{\text{ANEC}}(j,\jb) \quad \text{and} \quad M(\lambda_{O\Ob T},\Delta_O) \succeq 0\,,
}[eq:schematicANEC]
where $M$ is a matrix that depends linearly on the three-point function
coefficients $\lambda_{O\Ob T}$.  In a nonsupersymmetric theory, the
constraints on the three-point function coefficients generically admit a
solution.  Therefore, the first condition determines the bound on operator
dimensions.

In the presence of supersymmetry things can change significantly. First,
there exist certain multiplet shortening conditions, without a
nonsupersymmetric analog, that fix some of the three-point function coefficients
$\lambda_{O\Ob T}$, thus leaving less freedom to satisfy
\eqref{eq:schematicANEC}. Moreover, even without imposing any shortening
conditions, the ANEC must hold on any state $|\psi\rangle$ given by the
most general superposition of states in a super-multiplet---schematically
\eqn{
|\psi\rangle \sim \left(O +\alpha\lsp QO + \beta\lsp
\Qb O + \ldots \right) |0\rangle\,.
}[eq:super-state]
Computing the integral \eqref{eq:ANEC} on states \eqref{eq:super-state}
leads to more intricate constraints on the three-point function coefficients $\lambda_{O\Ob
T}$, which sometimes do not admit a solution. Intuitively, we then expect
that in the presence of supersymmetry a broader class of operators will
violate the ANEC and must thus be absent in any unitary local SCFT.

In the remainder of this section we briefly outline the logic behind our
computations and present our final results.  The rest of the paper
carefully goes through the details of our calculations.

\subsec{Strategy}
In this work we focus on superconformal multiplets $\mathcal O (x,\theta,\thetab)$
for which the lowest component field  $O$ transforms in
the $(\frac12j,0)$ irreducible representation of the Lorentz group. Our first goal is to
 determine the most general form of the three-point function in superspace among $\mathcal O$,
 its complex conjugate superfield, and the Ferrara--Zumino multiplet
 $\CJ$,
 which contains the stress-energy tensor:\footnote{In this section we only present
   schematic formulas. Details are given in the next sections.}
\eqn{
\langle \cOb (\mathbf z_1) \mathcal J (\mathbf z_2)\mathcal O (\mathbf z_3)  \rangle ,
\qquad \mathbf z_i = (x_i,\theta_i,\thetab_i)\,.
}[eq:schematic3pt]
In order to determine \eqref{eq:schematic3pt}, in Sec.~\ref{sec:Set-up} we
write the most general three-point function consistent with $\mathcal N=1$
superconformal invariance, complex conjugation, and conservation of the
Ferrara--Zumino multiplet. Next, we  fix certain combinations of the
three-point function coefficients entering \eqref{eq:schematic3pt} by
imposing the  Ward identities generated by the conserved currents
$J_R^\mu$, $T_{\mu\nu}$ and $S^\alpha_\mu$ in $\mathcal J$.  Although in
principle it should be possible to obtain a superspace version of the Ward
identities, along the lines of \cite{Osborn:1998qu}, in this work we impose
the constraints at the level of the individual primaries and
superdescendants.  More specifically, we find that once the $J_\mu^R$ and
$T_{\mu\nu}$ Ward identities are imposed in the three-point function
involving the superprimary $O$, all other ones we checked
follow.\footnote{More specifically, we checked the Ward identities for
$\langle (\Qb \Ob) J_R (Q O) \rangle$, $\langle (\Qb \Ob) T (Q O) \rangle$
and $\langle (\Qb \Ob) S O \rangle$. In principle there could be extra
relations that we did not take into account.}

In addition to the above, the three-point function \eqref{eq:schematic3pt}
could satisfy further constraints, associated to various shortening
conditions of the superconformal multiplet $\mathcal O$.  Following the
convention of \cite{Cordova:2016emh} we denote $\mathcal N=1$ multiplets as
$[\cX_L,\overbar{\cX}_R]$, where ${\cX_{L,R}}$ represents the action of the
charges $Q$ and $\Qb$, which give rise to long  ($L$), semi-short ($
{A_1},{A_2}$) or chiral ($ B$) multiplets.  We spell out the exact
definitions in \sectionname~\ref{sec:shortening}, together with the
corresponding unitarity bounds, and we compute the most general form of
\eqref{eq:schematic3pt} compatible with these constraints. The results are
summarized in the Tables in Appendix ~\ref{appendix:Ward}.

As a final step, we need to decompose the superspace three-point function
in the various $\theta$ components and extract the nonsupersymmetric
three-point functions of the superprimary $O$ and various primary
superdescendants.  This task is performed in
\sectionname~\ref{sec:expansion} and summarized in the Tables in Appendix
\ref{appendix:expansion-tables}.  Unfortunately the computations required
to perform this step become rapidly very complicated. In this work we only
pushed to the fourth order in $\theta_i$ or $\thetab_i$ and computed
three-point functions involving at most $T_{\mu\nu}$ and superdescendants
$Q O^\pm$ and $\Qb O$.\footnote{We remind that the action of a supercharge
produces in general two distinct primary superdescendants, which we
schematically denote with $\pm$, with equal dimension and R-charge but
transforming in different Lorentz representations. For operators in the
$(\frac12 j,0)$ representation only $\Qb O^+$ exists, so we will denote it
as $\Qb O$.}

After all these preparatory steps, we can impose the ANEC \eref{eq:ANEC} on
a general state of the form of \eref{eq:super-state}. Due to R-charge
conservation, only a few three-point functions are non vanishing.  In the
end we impose that\footnote{For certain short supermultiplets some of these
three-point functions vanish.}
\eqn{
\langle O | \mathcal E | O\rangle \geq 0 \,,\;\quad
\langle (\Qb O) | \mathcal E |  (\Qb O) \rangle \geq0 \;\quad \text{and}\quad \;
\begin{pmatrix}
 \langle  (Q O^+) | \mathcal E |   (Q O^+) \rangle &  \langle  (Q O^+) | \mathcal E |   (Q O^-) \rangle \\
\langle  (Q O^-) | \mathcal E |   (Q O^+) \rangle &  \langle  (Q O^-) | \mathcal E |   (Q O^-) \rangle
\end{pmatrix}\succeq 0\,.
}[eq:all-anec]
We should stress that the above conditions are a subset of all conditions
one can impose, since they do not include superdescendants of the form $Q^2
O$ or $Q\Qb O$ for example. Nevertheless, we find that in any unitary and
local SCFT superprimaries that transform in the $(\frac12 j,0)$
representation and satisfy the usual unitarity bounds do not necessarily
satisfy the conditions \eref{eq:all-anec}.

In \sectionname~\ref{sec:ANEC} we obtain closed-form expressions for all
the correlators appearing in \eref{eq:all-anec} as rational functions of
the spin $j$ and dimension $\Delta$.  Such formulas allow us to easily
compute bounds up to large values of $j$ and in some cases rigorously prove
bounds for any $j$.

Finally, we explore the  consequences of our analysis for theories with
extended supersymmetry. In \sectionname~\ref{sec:extendedSUSY} we consider
special $\mathcal N=2$ and $\mathcal N=4$ supermultiplets and decompose
them with respect to an $\mathcal N=1$ subalgebra. The ANEC constraints
presented in the next subsection are then recast as bounds on the $\mathcal
N=2,4$ superprimaries.

\subsec{Summary of results}
Let us first mention the results of our analysis for nonsupersymmetric CFTs, in the case of  a conformal primary with dimension
$\Delta$, transforming in $(\frac12 j, \frac12 \jb)$ representation, with $\jb=0,1$.
In \sectionname~\ref{sec:nonsusy} we show convincing evidence that the ANEC requires
\eqn{
  \Delta\geq \min\big( j, \tfrac{1}{15}(13\lsp j+42)\big)\,.
}[eq:nonsusyANEC]
For $\jb=0,1$  the above expression is stronger than the corresponding
unitarity bound for $j>2,6$, respectively. Although we don't have an
analytic proof, we checked \eref{eq:nonsusyANEC} up to $j=10^3$.

Next, we summarize the results of applying the ANEC inequality to
superconformal multiplets $\CO^{(j,0)}$.  We present them as bounds on the
quantum numbers $q,\bar q$, which are related to the dimension and the
R-charge of a given operator by the simple relations
\eqn{\Delta=q+\bar q\,, \qquad R = \tfrac23(q-\bar q)\,.}[]
We considered all possible shortening conditions classified
in~\cite{Cordova:2016emh} and we also follow their notation,\footnote{In a nutshell,
$L$ (resp. $\overbar L$) stand for long, $B$ (resp. $\overbar B$) for short or chiral,
 $A$ (resp. $\overbar A$) for semi-short with respect to the supercharge $Q$ (resp. $\Qb$).}
 which we briefly explain in
\sectionname~\ref{sec:shortening}.

\paragraph{All cases for $\boldsymbol{j=0}$} In this case the ANEC requires
only $q\geq 0$ and $\qb \geq 0$. Therefore, it is never stronger than the unitarity bound.

\paragraph{$\boldsymbol{[A_1,\overbar{B}]}$ for $\boldsymbol{j = 1}$} For these operators there are no free three-point function coefficients and the dimension and R-charge are fixed. It can be easily verified that the ANEC holds.

\paragraph{$\boldsymbol{[A_1,\overbar{B}]}$ for $\boldsymbol{j \geq 2}$}
As shown in \tablename~\ref{tab:final}, these operators do not admit a
three-point function with the Ferrara--Zumino multiplet consistent with all conditions. They are therefore absent in any local SCFT.\footnote{This conclusion does not require the ANEC.}

\paragraph{$\boldsymbol{[L,\overbar{B}]}$ for $\boldsymbol{j \geq 1}$}
With this shortening condition, corresponding to chiral operators, there
are no free three-point function coefficients. Therefore the ANEC for any
given $j$ is simply a system of inequalities on $q$ that can be solved
algebraically. The result is
\eqn{
\Delta=q \geq \tfrac32 j\,.
}[eq:chiralbound]
This is equivalent to the unitarity bound for $j = 1$ and it is stronger for all $j> 1$. This result is not in contradiction with already known Lagrangian constructions, which so far have only provided examples for $j=1$~\cite{Ceresole:1999zs,Cachazo:2002ry}. Also note that the bound is saturated by $j$ copies of a free $j=1$ superconformal chiral primary $\psi_i^\alpha$
\eqn{
\Psi^{\alpha_1\ldots \alpha_j} = \,:\lnsp\psi_1^{(\alpha_1}\cdots \psi_j^{\alpha_j)}\lnsp\lnsp:\,.
}[]
In $\mathcal N=2$ theories, the bound in \eref{eq:chiralbound} implies a
constraint on the dimension of the so called ``exotic chiral primaries''
discussed in \cite{Buican:2014qla}. In \sectionname~\ref{sec:N=2} we show that
\eqn{
\Delta_\text{exotic} \geq \tfrac{3}{2}j + 1\,.}[]

\begin{figure}[H]
\centering
\includegraphics{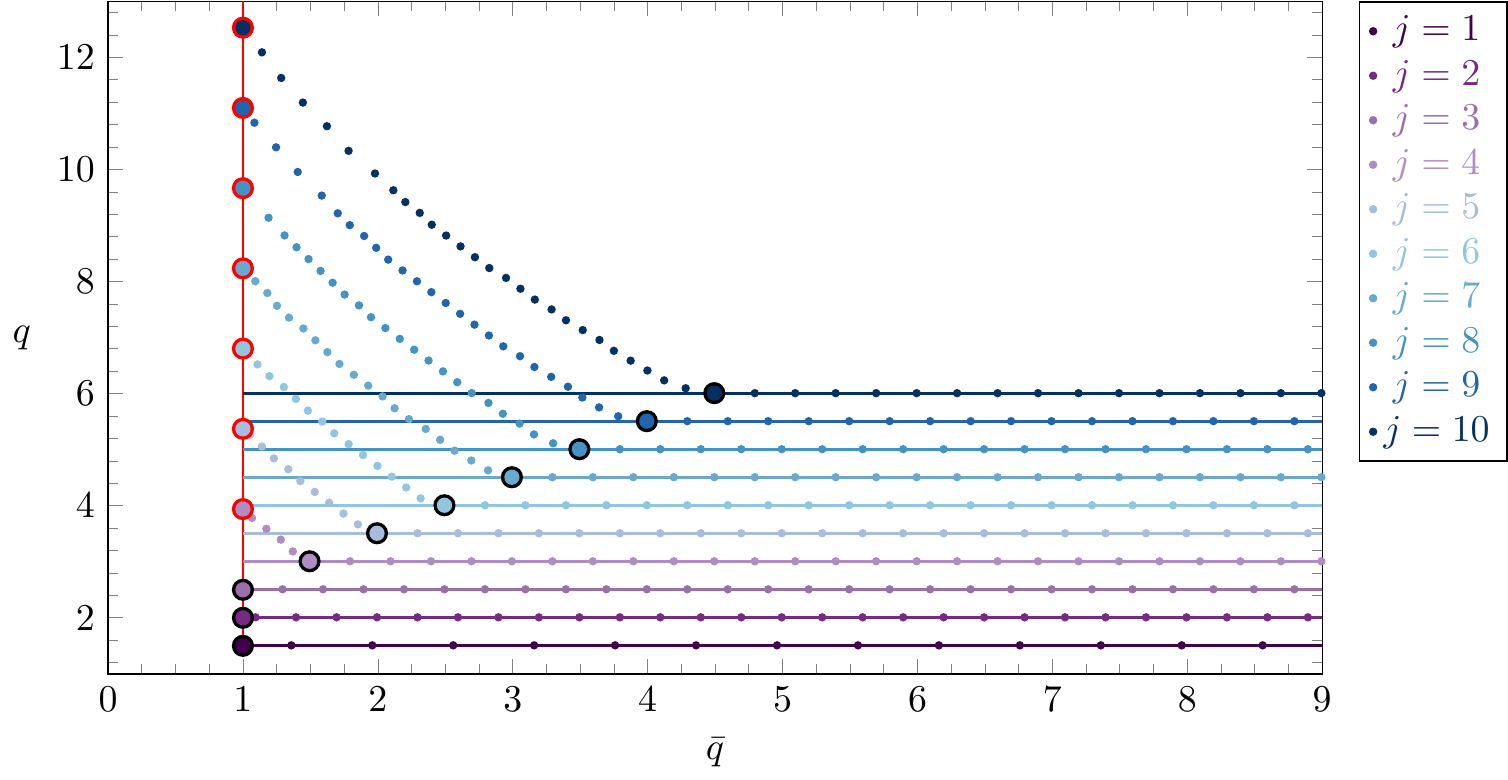}
\caption{Lower bounds on the conformal dimension as a result of the ANEC
for long multiples.  Each point is the result of a bisection algorithm done
with \href{https://github.com/davidsd/sdpb}{\texttt{sdpb}}~\cite{Simmons-Duffin:2015qma} (see \sectionname~\ref{ANECsemidef}). The solid lines are the unitarity bound: the red line is the bound on $\qb$ and the colored lines are the $j$-dependent bounds on $q$. The larger dots correspond to the points with shortening conditions $[L,\overbar{A}_2]$ (for the red circled dots) and $[A_1,\overbar{L}]$ (for the black circled dots).}\label{fig:LLb}
\end{figure}

\begin{figure}[H]
\centering
\includegraphics{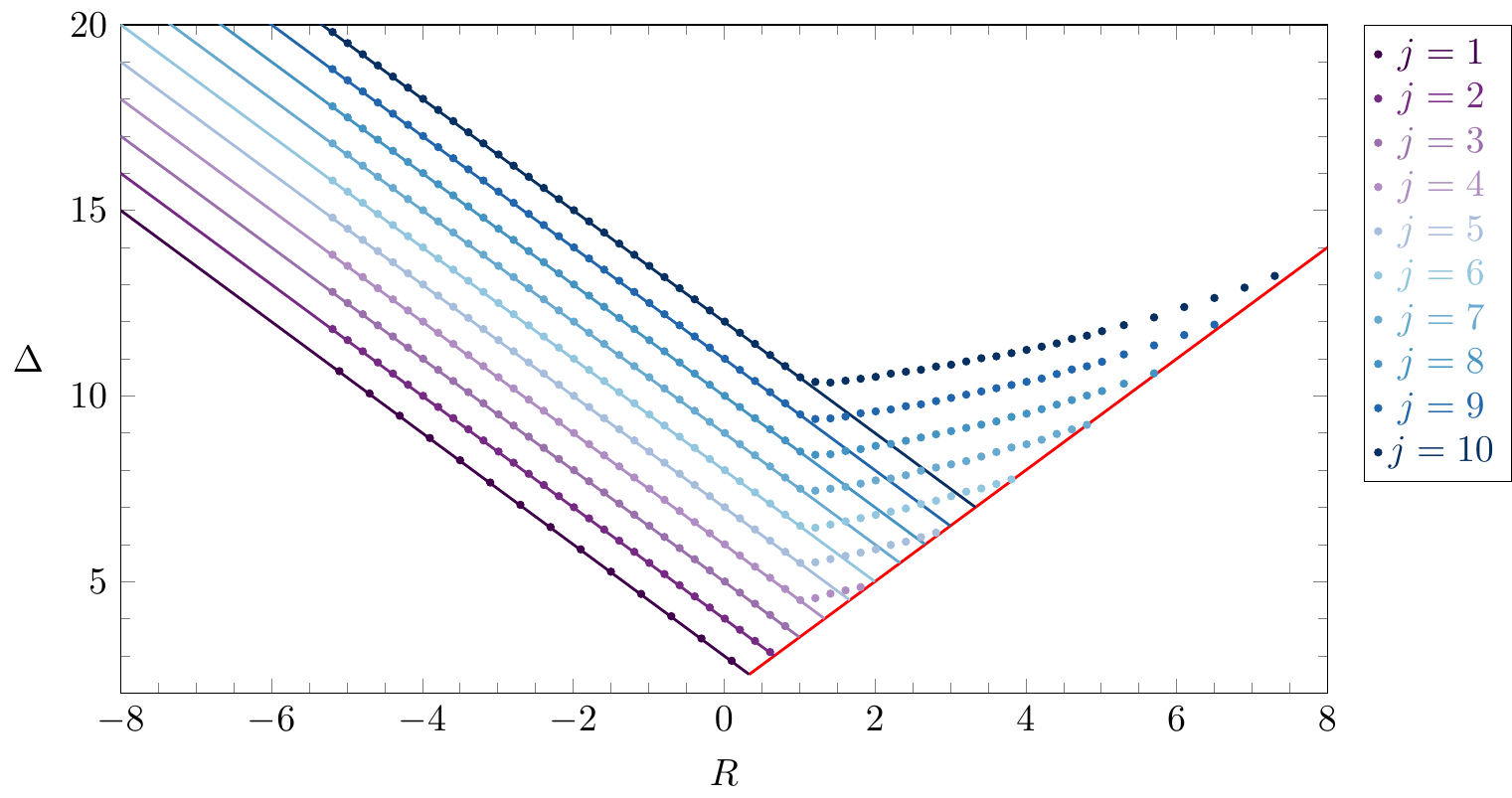}
\caption{Plot of the results in \figurename~\ref{fig:LLb} in the $(R,\Delta)$ plane.}\label{fig:LLbRDelta}
\end{figure}

\paragraph{$\boldsymbol{[L,\overbar{L}]}$ for $\boldsymbol{j \geq 1}$} In
this case there are two free parameters $q$ and $\qb$ and two undetermined
three-point function coefficients (one for $j=1$). For every value of $j$ we fixed $\qb$ and ran a bisection algorithm on $q$. The results are in \figurename~\ref{fig:LLb}. See also \figurename~\ref{fig:LLbRDelta} for a plot in the $(R,\Delta)$ plane.

\begin{figure}[H]
\centering
\includegraphics{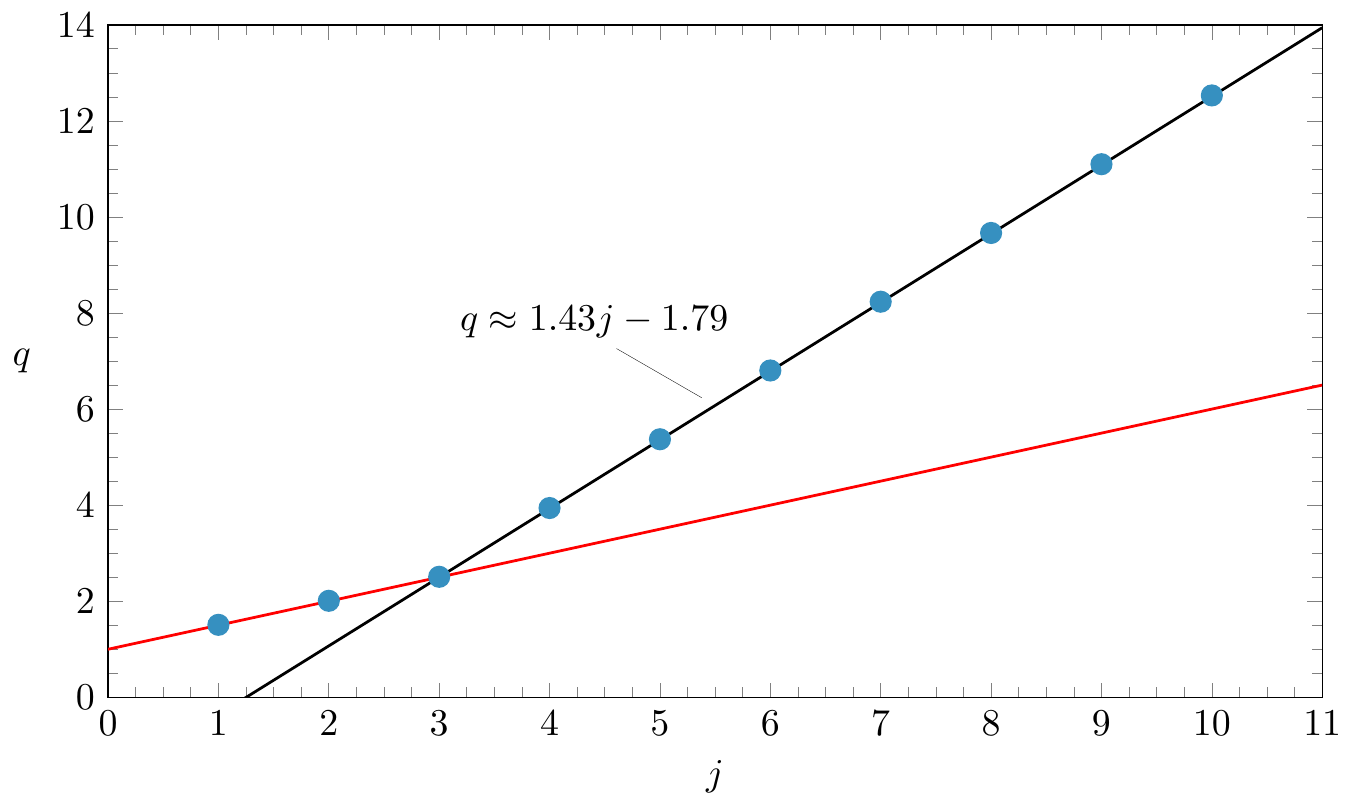}
\caption{Lower bounds on the conformal dimension as a result of the ANEC for $[L,\overbar{A}_2]$ multiplets. Each point is the result of a bisection algorithm done with \href{https://github.com/davidsd/sdpb}{\texttt{sdpb}}~\cite{Simmons-Duffin:2015qma} (see \sectionname~\ref{ANECsemidef}). The red line is the unitarity bound $q = \frac12 j + 1$. The operators for $j\leq 3$ that lie on the red line satisfy $[A_1,\overbar{A}_2]$.
}\label{fig:LAb2}
\end{figure}

\paragraph{$\boldsymbol{[L,\overbar{A}_2]}$ for $\boldsymbol{j \geq 1}$}
For this shortening condition the constraints are equivalent to
$[L,\overbar{L}]$ for $\qb = 1$. The results are given in
\figurename~\ref{fig:LAb2} and correspond to the red circled dots on
\figurename~\ref{fig:LLb}. The operators at the unitarity bound, which
satisfy the $[A_1,\overbar{A}_2]$ shortening, are not allowed for $j>3$
(see below). Therefore, for $j>3$ the ANEC provides a constraint strictly
stronger than unitarity.

\begin{figure}[H]
\centering
\includegraphics{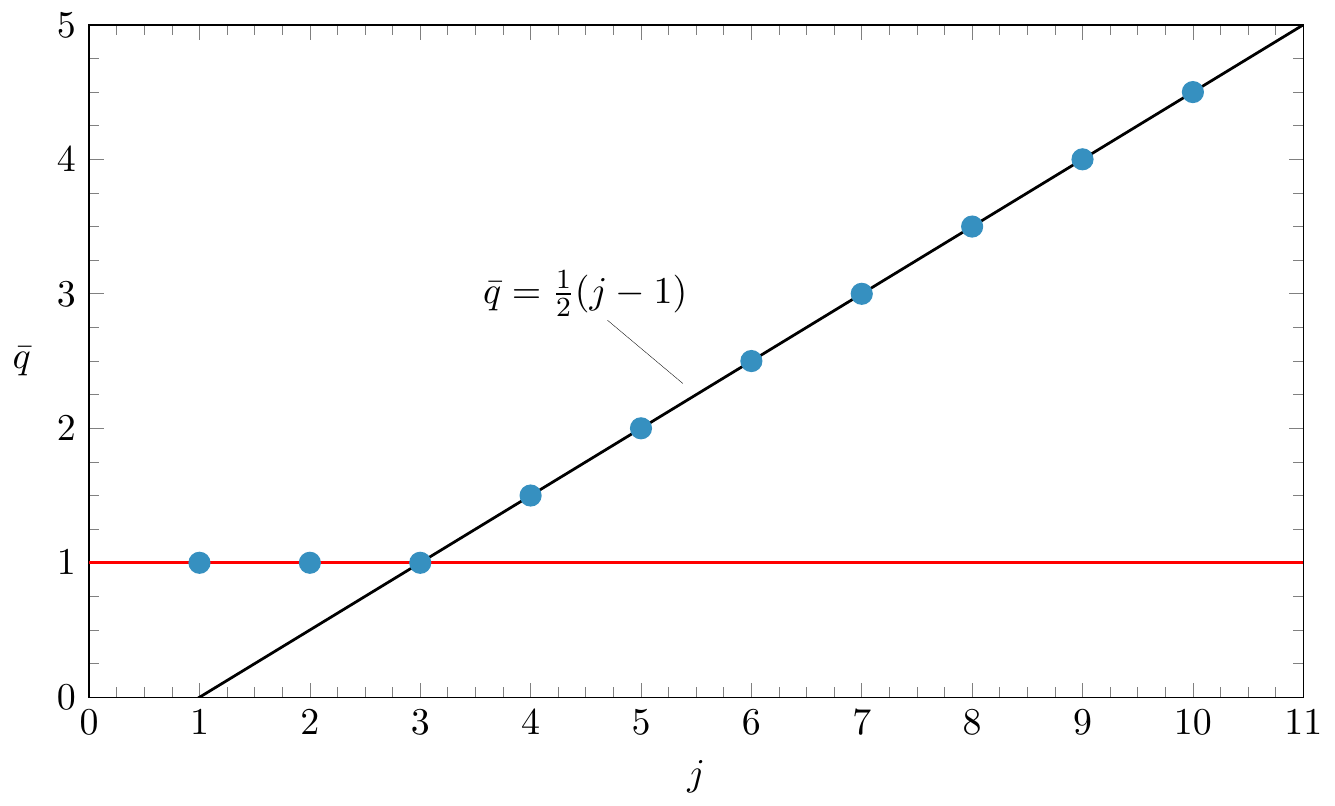}
\caption{Lower bounds on the conformal dimension as a result of the ANEC for $[A_1,\overbar{L}]$ multiplets. Each point is the result of a bisection algorithm done with \emph{Mathematica}. The operators for $j\leq 3$ that lie on the red line satisfy $[A_1,\overbar{A}_2]$.
}\label{fig:A1Lb}
\end{figure}

\paragraph{$\boldsymbol{[A_1,\overbar{L}]}$ for $\boldsymbol{j \geq 1}$}
Since for this case there is only one free three-point function coefficient
and one parameter, $\qb$, the system of inequalities is considerably simpler to solve. The results are given in \figurename~\ref{fig:A1Lb} and correspond to the black circled dots on \figurename~\ref{fig:LLb}. As before, for $j>3$, the ANEC is strictly stronger than unitarity.

\paragraph{$\boldsymbol{[A_1,\overbar{A}_2]}$ for $\boldsymbol{j \geq 1}$}
This condition admits solutions only for $j \leq 3$. In the edge case $j=3$ the ANEC inequalities fix the
only independent three-point function coefficient to
\eqn{
\cOJOF = -\frac{16}{\pi^2}\,.
}[eq:A1A2fixesB]
For all $j>3$ the ANEC admits no solution and thus such operators must be absent in any local SCFT.

\section{Setup}
\label{sec:Set-up}
Our object of study will be the three-point correlator in four dimensional
$\cN = 1$ superspace of a superconformal multiplet $\cO^{(j,0)}(\bfz)$,
its conjugate $\cOb^{(0,j)}(\bfz)$ and the Ferrara--Zumino multiplet
$\cJ^{(1,1)}(\bfz)$. The charges of $\cJ$ are $q_\cJ = \qb_\cJ =
\frac{3}{2}$, while those of $\cO$ and $\cOb$ are $q_\cO = \qb_\cOb = q$
and $\,q_\cOb = \qb_\cO = \qb$. The superscript $(j,\jb)$ refers to the
$\mathrm{SO}(3,1)$ representation,\footnote{Following standard conventions
we denote the irreducible representations of the Lorentz group by
$(\frac12 j,\frac12\jb)$, where $j$ is the number of undotted and
$\jb$ the number of dotted indices.} and will be henceforth dropped for brevity. The shorthand $\bfz$ denotes
\beq
\mathbf{z}_i = (z_i,\eta_i,\etab_i)\,,\quad \mbox{where}\quad z_i = (x_i, \theta_i,\thetab_i)\,.
\eeq
The polarizations $\eta_i,\,\etab_i$ are commuting spinors used to contract
all free indices as follows:
\beq
\cO^{(j,\jb)}(\bfz) = \eta^{\alpha_1}\cdots \eta^{\alpha_j}\etab^{\alphad_1}\cdots \etab^{\alphad_\jb} \cO_{\alpha_1\ldots\alpha_j\alphad_1\ldots\alphad_\jb}(z)\,.
\eeq
The tensor can be recovered by using spinor derivatives $\partial_\eta,\,\partial_\etab$ which obey
$\partial_{\eta^\alpha}\eta^\beta = \delta^\beta_\alpha$ and
$\partial_{\eta^\alpha}\eta_\beta = \epsilon_{\beta\alpha}$, and similarly
for the barred counterparts. We will follow the notation
of~\cite{Wess:1992cp} and the formalism introduced in~\cite{Osborn:1998qu}.

The most general three-point function involving $\cO^{(j,0)}(\bfz)$ can be written as
\eqn{
\big\langle\cOb(\bfz_1)\cJ(\bfz_2)\cO(\bfz_3)\big\rangle =
\frac{(\partial_{\chi_1}\lnsp\rmx_{3\bar{1}}\llsp\etab_1)^j\,\eta_2\rmx_{2\bar{3}}\partial_{\chib_2}\,\partial_{\chi_2}\lnsp\rmx_{3\bar{2}}\llsp\etab_2}{{x_{\bar{1}3}}{\!}^{2q+j}\,{x_{\bar{3}1}}{\!}^{2\qb}\,{x_{\bar{3}2}}{\!}^4\, {x_{\bar{2}3}}{\!}^4}\,t(Z_3;\chi_1,\chi_2,\chib_2,\eta_3)\,,
}[eq:threepfSuperspace]
where $Z_3 = (X_3,\Theta_3,\Thetab_3)$ will be defined shortly and
$\chi_i,\,\chib_i$ are auxiliary spinor polarizations.\footnote{They are
denoted with a different letter than $\eta$ only to emphasize the fact that
they are eventually removed by the derivatives in the numerator.}
The function $t$ can be expressed as a linear combination of ten tensor structures, but the coefficients multiplying them are not arbitrary as they are constrained by reality of the correlator, conservation of $\cJ$, eventual shortening conditions on $\cO$ and the Ward identities for the R-symmetry and the conformal group. We will analyze all these constraints in the next section. Let us now briefly describe all the fundamental building blocks of such tensor structures. They are functions of the supersymmetric interval $\rmx_{i\jb}$ defined as
\beq
(\rmx_{i\jb})_{\alpha\alphad} = -\epsilon_{\alpha\beta}\epsilon_{\alphad\betad} (\tilde{\rmx}_{\jb i})^{\betad\beta} = -\sigma^\mu_{\alpha\alphad} (x_{\jb i})_\mu = (\rmx_{ij})_{\alpha\alphad}-2i\lsp \theta_{i\alpha}\thetab_{i\alphad} - 2i\lsp \theta_{j\alpha}\thetab_{j\alphad} +4i\lsp \theta_{i\alpha}\thetab_{j\alphad}\,,
\eeq
with $x_{ij}=x_i-x_j$, ${x_{\ib j}}^a =({x_{\ib j}}^2)^{a/2} $ and of the
Grassmann intervals $\theta_{ij} = \theta_i-\theta_j$, $\thetab_{ij} = \thetab_i-\thetab_j$. We can use these objects to define
\eqna{
&\mathrm{X}_3 = \frac{\rmx_{3\bar{1}}\tilde{\rmx}_{\bar{1}2}\rmx_{2\bar{3}}}{{x_{\bar{1}3}}{\!}^2{x_{\bar{3}2}}{\!}^2}\,,\quad
\overbarUp{\mathrm{X}}_3 = - \frac{\rmx_{3\bar{2}}\tilde{\rmx}_{\bar{2}1}\rmx_{1\bar{3}}}{{x_{\bar{3}1}}{\!}^2{x_{\bar{2}3}}{\!}^2} = \mathrm{X}_3 ^\dagger\,,\quad \\
&
\Theta_3 = i \left(\frac{\rmx_{3\bar{1}}\thetab_{31}}{{x_{\bar{1}3}}{\!}^2}-\frac{\rmx_{3\bar{2}}\thetab_{32}}{{x_{\bar{2}3}}{\!}^2}\right)\,,\quad \Thetab_3 = i \left(\frac{\theta_{31}\rmx_{1\bar{3}}}{{x_{\bar{3}1}}{\!}^2}-\frac{\theta_{32}\rmx_{2\bar{3}}}{{x_{\bar{3}2}}{\!}^2}\right)=\Theta_3^\dagger \,.
}[]
Similar objects $\mathrm{X}_i,\,\Theta_i,\,\Thetab_i$, $i=1,2$, can be defined by a cyclic permutation of the points. We will further define
\beq
\mathrm{U}_3 = \lifrac12(\mathrm{X}_3+\overbarUp{\mathrm{X}}_3)\,.
\eeq
Also, note that $\mathrm{X}_3-\overbarUp{\mathrm{X}}_3 = 4i\lsp
\Theta_3\Thetab_3$. We can then form fully contracted monomials of the
quantities defined above to obtain the building blocks of the tensor
structures in $t$. A complete list is
\eqn{
\begin{gathered}
  \Uc{i}{\jmath} = \frac{\eta_i\lnsp\mathrm{U}\etab_j}{|U|} \,,\quad \Uc{\Theta}{\Theta} = \frac{\Theta\mathrm{U}\Thetab}{U^2}\,,  \quad   \Ec{i}{j} = \eta_i \eta_j\,, \quad \Ecb{\imath}{\jmath} = \bar{\eta}_i \bar{\eta}_j\,,\quad [\Theta^2]=\frac{\Theta^2}{|U|}\,,\\ [\Thetab^2]=\frac{\Thetab^2}{|U|}\,,\quad
  \Ec{\Theta}{j} = \frac{\Theta \eta_j}{|U|^{1/2}} \,,\quad
    \Ecb{\Theta}{\jmath} = \frac{\Thetab \etab_j}{|U|^{1/2}} \,,\quad
    \Uc{j}{\Theta} = \frac{\eta_i\lnsp\mathrm{U}\Thetab}{|U|^{3/2}}\,, \quad \Uc{\Theta}{\jmath} = \frac{\Theta\lnsp\mathrm{U}\etab_j}{|U|^{3/2}}\,.
\end{gathered}}[Shortcuts]

Other than the physical constraints mentioned before, that will be addressed later, $t$ must satisfy certain homogeneity properties, which can be summarized as
\eqn{
t(\lambda \bar{\lambda} X,\lambda \Theta,\bar{\lambda}\Thetab;\kappa\eta_1,\mu\eta_2,\bar{\mu}\etab_2,\bar{\kappa}\eta_3) =
(\lambda\bar{\lambda})^{-3}
(\kappa\bar{\kappa})^j \mu\bar{\mu}\, t(X,\Theta,\Thetab;\eta_{i},\etab_{i})\,.
}[eq:homogeneityEta]
All possible tensor structures are built out of the blocks in \Shortcuts times a factor $U^{-3}$ to take care of the $\lambda\bar{\lambda}$ scaling. Not all combinations will be independent due to several relations termed \emph{Schouten identities} which stem from the vanishing of $\epsilon^{\alpha[\beta}\epsilon^{\gamma\delta]}$ and the corresponding tensor with dotted indices. We will now produce a list of ten tensor structures that we have explicitly checked to be linearly independent. We can then claim it is a basis because it agrees with the expected number of structures obtained with a group theoretic formula introduced in \cite{Manenti:2018xns} as a superspace generalization of \cite{Kravchuk:2016qvl}.\par
As already mentioned, $t$ can be written as a linear combination
\eqn{
t(Z;\eta_1,\eta_2,\etab_2,\eta_3) = \frac{1}{U^3}\sum_{k=1}^{10} \cOJOi{k}\,\susyT{k}(Z;\eta_1,\eta_2,\etab_2,\eta_3)\,.
}[]
The explicit expressions for the $\susyT{k}$'s are
\eqn{
\begin{aligned}
  \susyT{1} &= i\Uc22\Ec13^j &  \susyT{6} &= \Ec12 \Uc12 \Ec{\Theta}3\Uc3{\Theta}\Ec13^{j-2}\\
\susyT{2} &= i\Ec12\Uc32\Ec13^{j-1} &  \susyT{7} &= \Ec12\Ecb{\Theta}2\Ec{\Theta}3\Ec13^{j-1}\\
\susyT{3} &= \Uc3{\Theta}\Ec{\Theta}2\Uc12 \Ec13^{j-1} \qquad & \susyT{8}  &= \Ec12\Uc32\Uc{\Theta}{\Theta}\Ec13^{j-1}\\
\susyT{4} &= \Ec{\Theta}2\Ecb{\Theta}2\Ec13^j & \susyT{9} &= i\lsp [\Theta^2]\lsp[\Thetab^2] \Uc22 \Ec13^j\\
\susyT{5} &= \Uc22\Uc{\Theta}{\Theta}\Ec13^j & \susyT{10} &= i\lsp [\Theta^2]\lsp[\Thetab^2] \Ec12\Uc32 \Ec13^{j-1}\,.\\
\end{aligned}
}[]
The factors of $i$ are introduced for later convenience. If $j=1$ then $\susyT{6}$ is not present and if $j=0$ then $\susyT{2,3,6,7,8 ,10}$ are not present.

\section{Constraints on the supersymmetric three-point correlator}
\subsection{Conservation}
The superconformal multiplet $\cJ(\bfz)$ contains the R-symmetry current,
the supersymmetry current and the stress-energy tensor. As a consequence, it satisfies a shortening condition (see e.g. \cite{Komargodski:2010rb}) which can be expressed in superspace as
\eqn{
D^\alpha\cJ_{\alpha\alphad}(\bfz) = \overbar{D}^\alphad\cJ_{\alpha\alphad}(\bfz) = 0\,,
}[]
with $D$ and $\overbar{D}$ the superspace derivatives. In this subsection
we will explore the consequences of this constraint on the correlator at
separated points. In \sectionname~\ref{sec:WI} we will study the contact
terms instead. At separated points the prefactor in \eqref{eq:threepfSuperspace} commutes with the conservation differential operators acting on $\bfz_2$,\footnote{Due to $D_2^\alpha (\rmx_{2 \bar{3}})_{\alpha\alphad}/x_{\bar{3}2}^4=\overbar{D}_2^\alphad (\rmx_{3 \bar{2}})_{\alpha\alphad}/x_{\bar{2}3}^4 = 0$ when $x_{23} \neq 0$.} thus we can express the conservation condition as an equation involving only $t$ and the variable $Z$:
\eqn{
\partial_{\eta_2}\cD\,t(Z;\eta_i,\etab_i) = \partial_{\etab_2}\overbar{\cD}\,t(Z;\eta_i,\etab_i) = 0\,,
}[eq:consEq]
where
\eqn{
\cD_{\alpha} = \frac{\partial}{\partial \Theta^\alpha} -  i  \sigma^\mu_{\alpha\alphad}\lsp\Thetab^\alphad \frac{\partial}{\partial U^\mu}\,,\qquad
\overbar{\cD}_{\alphad} = \frac{\partial}{\partial \Thetab^\alphad}+ i  \Theta^\alpha\sigma^\mu_{\alpha\alphad} \frac{\partial}{\partial U^\mu}\,.
}[]
Equation \eqref{eq:consEq} imposes the following linear constraints for
general $j>1$:
\eqn{
\cOJOE = - \cOJOC-2\lsp\cOJOD\,,\quad  \cOJOG = 2\lsp\cOJOB-\cOJOC -\cOJOF\,,\quad \cOJOH = -4\lsp\cOJOB+2\lsp\cOJOC +\cOJOF\,,\quad \cOJOI=\cOJOJ=0\,.
}[eq:consCond]
When $j=1$ it suffices to set $\cOJOF$ to zero and when $j=0$ one simply has
\eqn{
\cOJOE = -2\lsp \cOJOD\,,\qquad \cOJOI = 0\,.
}[]

\subsection{Reality}
Since $\cO$ and $\cOb = \cO^\dagger$ are conjugated to each other and $\cJ$
is hermitian, the correlator under study must be real. Concretely, we want to impose that
\eqn{
\Vev{\COb(\bfz_1)\CJ(\bfz_2)\CO(\bfz_3)}^* = \Vev{\COb(\bfz_3)\CJ(\bfz_2)\CO(\bfz_1)}\,,
}[eq:reality]
namely that taking the complex conjugation is the same as swapping points
$1$ and $3$. The prefactor in \eqref{eq:threepfSuperspace} is not invariant
under this transformation, moreover the exchange $1\leftrightarrow 3$ does
not act nicely on $Z_3$. This means that we cannot translate the reality
condition into a constraint for $t$ right away.\footnote{This is obviously
a consequence of our parametrization. In the ordering $\vev{\cO \cOb \cJ}$
the reality condition can be solved easily. On the other hand we would lose
the fact that the conservation operator commutes with the prefactor thus
making conservation much harder to impose.} We proceed, then, to expand the definition of \eqref{eq:reality}
\eqna{
&\frac{(-\eta_1\rmx_{1\bar{3}}\partial_{\chib_1})^j\,\eta_2\rmx_{2\bar{3}}\partial_{\chib_2}\,\partial_{\chi_2}\lnsp\rmx_{3\bar{2}}\llsp\etab_2}{{x_{\bar{3}1}}{\!}^{2q+j}\,{x_{\bar{1}3}}{\!}^{2\qb}\,{x_{\bar{2}3}}{\!}^4\, {x_{\bar{3}2}}{\!}^4}\,t(Z_3;\chi_1,\chi_2,\chib_2,\eta_3)^*=\\
&\hspace{3cm}=
\frac{(\partial_{\chi_1}\lnsp\rmx_{1\bar{3}}\llsp\etab_3)^j\,\eta_2\rmx_{2\bar{1}}\partial_{\chib_2}\,\partial_{\chi_2}\lnsp\rmx_{1\bar{2}}\llsp\etab_2}{{x_{\bar{3}1}}{\!}^{2q+j}\,{x_{\bar{1}3}}{\!}^{2\qb}\,{x_{\bar{1}2}}{\!}^4\, {x_{\bar{2}1}}{\!}^4}\,t(\Zb_1;\chi_1,\chi_2,\chib_2,\eta_1)\,,
}[eq:realityExpanded]
where we defined, $\Zb_1 = (-\Xb_1,-\Theta_1,-\Thetab_1)$. The prefactor
appearing in the above expression can be recast in terms of the
supersymmetric inversion tensor $I^{i\ib}$ introduced in
\cite{Osborn:1998qu}, which we review in \appendixname~\ref{app:tensorsI}.
The indices $i$ (resp.\ $\ib$) collectively denote $j$ symmetrized
$\alpha$ (resp.\ $(\alphad)$) indices and $\mu$ is an ordinary Lorentz vector index. In this notation \eqref{eq:realityExpanded} reads
\eqn{
\frac{{x_{\bar{2}1}}{\!}^3\, {x_{\bar{1}2}}{\!}^3}{{x_{\bar{2}3}}{\!}^3\, {x_{\bar{3}2}}{\!}^3}I^{i_1\ib_1}(x_{1\bar{3}})\,I_{\mu\nu}(x_{2\bar{3}},x_{\bar{2}3})\lsp(t^*)^{\phantom{\ib_1}\nu}_{\ib_1\phantom{\nu\,}\ib_3}(Z_3) = \bar{I}_{\ib_3 i_3}(x_{\bar{3}1})I_{\mu\nu}(x_{2\bar{1}},x_{\bar{2}1})\,t^{i_3\nu \lsp i_1}(\Zb_1)\,.
}[]
Contracting both sides of this expression with
$\bar{I}^{\sigma\lambda}(\Xb_1,X_1)I^{i_4\ib_3}(x_{1\bar{3}})\,I_{\lambda\rho}(x_{1\bar{3}},x_{\bar{1}3})\,\bar{I}^{\rho\mu}(x_{\bar{3}2},x_{3\bar{2}})$
and using the various identities listed in \appendixname~\ref{app:tensorsI} we get to the final expression
\eqn{
I^{i_1\ib_1}(\Xb_1)I^{i_4\ib_3}(\Xb_1)\,(t^*)^{\phantom{\ib_1}\sigma}_{\ib_1\phantom{\mu\,}\ib_3}(Z_1) =t^{i_4\sigma \lsp i_1}(\Zb_1)\,,
}[]
which, in index-free form, reads\footnote{By $(t^*)(Z;\ldots)$ we mean: first apply the complex conjugation to $t(Z;\eta_1,\eta_2,\etab_2,\eta_3)$, then replace $\etab_{1(3)}$ with $\overbarUp{\mathrm{X}}\llsp\eta_{1(3)}$.}
\eqn{
(-1)^j\Xb^{-2j}\lsp(t^*)(Z;\overbarUp{\mathrm{X}}\llsp\eta_1,\eta_2,\etab_2,\overbarUp{\mathrm{X}}\llsp\eta_3) = t(\Zb;\eta_3,\eta_2,\etab_2,\eta_1)\,.
}[eq:realityEqANEC]
We can then solve this equation much more easily. In doing so we find the
following linear constraints for even $j>1$:
\eqna{
&\cOJOA^* = \cOJOA\,,\quad
\cOJOB^* = \cOJOB\,,\quad
\cOJOC^* = 2\lsp\cOJOB-\cOJOF-\cOJOG\,,\quad
\cOJOD^* = -2\lsp\cOJOB+\cOJOC+\cOJOD+\cOJOF+\cOJOG\,,\\
& \cOJOE^* = \cOJOE\,,\quad
\cOJOF^*=\cOJOF\,,\quad
\cOJOG^* = 2\lsp \cOJOB -\cOJOC-\cOJOF\,,\quad
\cOJOH^*=\cOJOH\,,\\
& \cOJOI^* = \cOJOB - \lifrac12 (\cOJOC+\cOJOF+\cOJOG)+ \cOJOI\,,\quad
\cOJOJ^* = -2\lsp \cOJOB + \cOJOC+\cOJOF+\cOJOG+\cOJOJ\,.
}[]
If $j$ is odd the equations are obtained by adding an overall minus sign on the right hand side. If $j=1$ it is sufficient to set $\cOJOF = \cOJOF^*=0$. For $j=0$ instead one has simply
\eqn{
\cOJOA^* = \cOJOA\,,\qquad
\cOJOD^* = \cOJOD\,,\qquad
\cOJOE^* = \cOJOE\,,\qquad
\cOJOI^* = \cOJOI\,.
}[]
Combined with conservation \eqref{eq:consCond}, these equations imply that
the remaining $\cOJOi{k}$ are purely real (resp.\ imaginary) if $j$ is even
(resp.\ odd).

\subsection{Ward identities}\label{sec:WI}

There are in principle two ways to impose the Ward identities: one could
apply them directly in superspace with the formalism
of~\cite{Osborn:1998qu}, or alternatively one could expand the correlator
in components and apply the nonsupersymmetric Ward identity to each
triplet of superdescendants. Since we already need the three-point function
in components to make contact with the ANEC and since nonsupersymmetric
Ward identities are much easier to compute, we opted for the second
approach. We did not explore all possible combinations of superdescendants
but we observed that after imposing the identities for $\langle \Ob
JO\rangle$ and $\langle \Ob T O\rangle$, all other choices of
superdescendants that we investigated were not yielding any new
constraints. By $O,\Ob$ we mean the lowest component of $\cO,\cOb$, while
the R-current $J_\mu$ and the stress-energy tensor $T_{\mu\nu}$ are, respectively, the lowest component and the $Q\Qb$ component of $\cJ$. We will also denote the supersymmetry currents as $S^\mu_\alpha$ and $\overbar{S}^\mu_\alphad$. They are, respectively, the $Q$ and the $\Qb$ components of $\cJ$. \par
For nonsupersymmetric three-point functions we use the conventions of~\cite{Cuomo:2017wme}.\footnote{We used their \emph{Mathematica} package \href{https://gitlab.com/bootstrapcollaboration/CFTs4D}{\texttt{CFTs4D}} to generate the tensor structures.} A three-point function $\mathsf{t}$ can be expressed as a prefactor multiplying a linear combination of tensor structures,
\eqn{
\mathsf{t}_{O_1O_2O_3}(x_{1,2,3},\eta_{1,2,3},\etab_{1,2,3}) \equiv \langle \mbox{$\prod_{i=1}^3$} O_i(x_i,\eta_i,\etab_i)\rangle = \cK \sum_k \lambda_k\,\nonsusyT{k}(x_{1,2,3},\eta_{1,2,3},\etab_{1,2,3}) \,,
}[]
where $\cK$ is of the form $\cK = \prod_{j>i} {x_{ij}}^{\delta_{ij}}$,
$\delta_{ij}$ being linear functions of the dimensions and spins of the
operators in the three-point function. The tensor structures $\nonsusyT{k}$
can be built out of the following invariants:
\eqn{
\invI{i}{j}\,,\qquad\invJ{i}{j}{k}\,,\qquad \invK{i}{j}{k}\,,\qquad \invKb{i}{j}{k}\,.
}
We refer the reader to \cite[Appendix D]{Cuomo:2017wme} for their
definition. For the two-point function we use the convention
\eqn{
\mathsf{n}_{\Ob O}(x_{12},\eta_{1,2},\etab_{1,2}) \equiv \langle \Ob(x_1,\eta_1,\etab_1)O(x_2,\eta_2,\etab_2)\rangle = c_O\,i^{j+\jb}\,\frac{(\eta_2\llsp\rmx_{12}\etab_1)^j(\eta_1\lnsp\llsp\rmx_{12}\etab_2)^\jb}{{x_{12}}^{2\Delta + j+\jb}}\,,
}[]
assuming $O$ has spin $(\frac12 j,\frac12\jb)$. In a unitary theory $c_O > 0$. The
coefficient $c_O$ is usually set to $1$ by normalizing the operator in the
two-point function, but here we do not do this rescaling of operator
because in the supersymmetric case the relative normalizations of the operators in the same superconformal multiplet are fixed. We will assume that the superconformal primary is normalized to $c_O = 1$ and use the results of \cite{Li:2014gpa} to fix the normalization of its superdescendants.

\paragraph{R-current} Let us start our analysis with the Ward identity for
the $U(1)_R$ symmetry. The three-point function under study is
$\mathsf{t}_{\Ob JO}$, where $O$ is any operator with charge $r$ under
$U(1)_R$ and $\Ob$ its conjugate (with charge $-r$). Now consider a
codimension-one surface $\Sigma$ enclosing $x_2$ and $x_3$ but not $x_1$. The Ward identity states
\eqn{
\frac{i}{2} \int_{\Sigma}\di \Omega(x_{23}) \,x_{23}^2
\,\partial_{\eta_2}\rmx_{23}\partial_{\etab_2}\,\mathsf{t}_{\Ob
JO}(x_i;\eta_i,\etab_i)  = \mathfrak{N}\,r \,\mathsf{n}_{\Ob O}(x_{13},\eta_{1,3},\etab_{1,3}) \,.
}[]
The factor $i/2$ on the left hand side comes from the $-1/2$ obtained from
$x^\mu J_\mu = -\frac12	\tilde{\rmx}^{\alphad\alpha}J_{\alpha\alphad}$ and
a $-i$ from the Wick rotation. Indeed the integral in the above equation is
in Euclidean signature and the right prescription for the Wick rotation is
the one that keeps the operators radially ordered as indicated, namely if
$x_i^0 = -i \tau_i$, then $\tau_1>\tau_2>\tau_3$. The constant
$\mathfrak{N}$ is a normalization for the multiplet $\cJ$. In order to
match the conventions of \cite{Cordova:2017dhq} we must set
$\mathfrak{N}=2$. Since this integral depends only topologically on the
points we can evaluate it in the simplified limit $x_1\to\infty,\,x_{23}
\to 0$. The results for an operator $O$ of spin $(\frac12 j,0)$ are summarized in
\tablename~\ref{tab:WIObJO} and those for $O$ of spin $(\frac12 j,\frac12)$ in
\tablename~\ref{tab:WIQObJQbO}.\footnote{The results showed in these tables
and the subsequent ones already assume the normalization $\mathfrak{N} =2$.}
\appendset{\begin{table}[!ht]
\centering
\begin{tabularx}{\textwidth}{XlllllX}
\hline
&$\cObJO{i}$ & Structure & $j>1$ & $j=1$ & $j=0$ &\\ \hline\hline
&$\cObJO{1}$ &
\tableeqn{\invJ132\lsp(\invI13)^j} &
\tableeqn{-\frac12\lsp\cObJO2 + \frac{2i^{j+1}(q-\qb)}{3\pi^2}} &
\tableeqn{-\frac12\lsp\cObJO2 - \frac{2\lsp(q-\qb)}{3\pi^2}} &
\tableeqn{\frac{2i\lsp(q-\qb)}{3\pi^2}}
\\\hline
&$\cObJO{2}$ &
\tableeqn{\invI23\lsp(\invI13)^{j-1}} &
\tableeqn{\cObJO{2}} &
\tableeqn{\cObJO{2}} &
\hfill\tableeqn{\diagup}\hfill\lsp\\
\hline
\end{tabularx}
\caption{Ward identities of the R-current for the correlator $\langle\Ob J
O\rangle$ when $O$ has spin $(\frac12 j,0)$.}
\label{tab:WIObJO}
\end{table}}{wi}

\appendset{\begin{table}[!ht]
\centering
\begin{tabularx}{\textwidth}{XlllllX}
\hline
&\hspace{-1ex}$\cQObJQbO{i}$ & Structure & $j>1$ & $j=1$ & $j=0$ &\\ \hline\hline
&\hspace{-1ex}$\cQObJQbO1$
& \tableeqn{\invJ231\lsp\invI23\lsp\invI32\lsp(\invI13)^{j-1}}
& \tableeqn{\cQObJQbO1}
&
&\hfill$\diagup$\hfill\lsp
\\\hline
&\hspace{-1ex}$\cQObJQbO2$
& \tableeqn{\invJ132\lsp\invI31\lsp(\invI13)^j}
& \tableeqn{\cQObJQbO2}
&
&
\\\hline
&\hspace{-1ex}$\cQObJQbO3$
& \tableeqn{\invI21\lsp\invI32\lsp(\invI13)^j}
& \tableeqn{\begin{aligned}
           &2\lsp\cQObJQbO2+\cQObJQbO5-\frac12(\cQObJQbO1+\cQObJQbO6+\cQObJQbO4)\\
           &-\frac{4i^j\lsp(3+2(q-\qb))}{3\pi^2}
           \end{aligned}}
&
&
\\\hline
&\hspace{-1ex}$\cQObJQbO4$
& \tableeqn{\invJ231\lsp\invJ123\lsp\invI12\lsp\invI23\lsp(\invI13)^{j-2}}
& \tableeqn{\cQObJQbO4}
& \hfill$\diagup$\hfill\lsp
&\hfill$\diagup$\hfill\lsp
\\\hline
&\hspace{-1ex}$\cQObJQbO5$
& \tableeqn{\invI12\lsp\invI23\lsp\invI31\lsp(\invI13)^{j-1}}
& \tableeqn{\cQObJQbO5}
&
&\hfill$\diagup$\hfill\lsp
\\\hline
&\hspace{-1ex}$\cQObJQbO6$
& \tableeqn{\invJ123\lsp\invI12\lsp\invI21(\invI13)^{j-1}}
& \tableeqn{\cQObJQbO6}
&
&\hfill$\diagup$\hfill\lsp
\\\hline
\end{tabularx}
\caption{Ward identities of the R-current for the correlator $\langle \Ob'
J O'\rangle$ when $O'$ has spin $(\frac12 j,\frac12)$, R-charge $\tfrac23(q-\qb)+1$ and is assumed to be unit normalized. If $O' = \Qb O$ the terms not proportional to $\cQObJQbO{k}$ must be rescaled by $c_{(\Qb O)}$. The unbarred entries in the $j=1,0$ columns are obtained by setting the absent coefficients to zero.}
\label{tab:WIQObJQbO}
\end{table}}{wi}

\paragraph{Stress-energy tensor} We proceed by considering the Ward identities for the conformal group. To each conformal Killing vector $\varepsilon^a_\mu$ is associated a possibly independent identity. It is sufficient to impose only $\varepsilon_\mu = x_\mu$ (dilatations) and $\varepsilon_\mu^\nu = \delta^\nu_\mu$ (translations).\footnote{Following~\cite[Appendix B]{Karateev:2019pvw} the independent constraints given by the Ward identities are as many as the number of singlets in
\[
  \rho_O\otimes\rho^\dagger_{\Ob}\otimes(\bullet \oplus
  (1,0)\oplus(0,1))\,,
\]
$\rho_O$ representing the Lorentz representation of $O$ and $\bullet$ the
singlet. For $\rho_O = (\frac12 j,0)$ the tensor product contains two
singlets (one if $j=0$) and for $\rho = (\frac12j,\frac12)$ it contains three singlets (two if $j=0$). The equations \eqref{eq:WIstressT} yield the exact same number of independent constraints.} Dilatations and translations imply respectively the  identities
\eqna{
-\frac{i}{8}\int_\Sigma\di \Omega(x_{23})\,x_{23}{\!}^2\,
\partial_{\eta_2}\rmx_{23}\partial_{\etab_2}\lsp\partial_{\eta_2}\rmx_{2}\partial_{\etab_2}\lsp
\mathsf{t}_{\Ob TO}(x_i,\eta_i,\etab_i) &= -i\lsp\mathfrak{N}\lsp\big(\Delta+x_3\cdot\partial_3\big) \,\mathsf{n}_{\Ob O}(x_{13},\eta_{1,3},\etab_{1,3})\,,
\\
-\frac{i}{8}\int_\Sigma\di \Omega(x_{23})\,x_{23}{\!}^2\,
\partial_{\eta_2}\rmx_{23}\partial_{\etab_2}\,\partial_{\eta_2}\mathrm{y}\partial_{\etab_2}\,
\mathsf{t}_{\Ob TO}(x_i,\eta_i,\etab_i) &= -i\lsp\mathfrak{N}\, y\cdot\partial_3\,\mathsf{n}_{\Ob O}(x_{13},\eta_{1,3},\etab_{1,3})\,,
}[eq:WIstressT]
where $y^\mu$ is an arbitrary vector used to contract the free index of the
translation Killing vector. The operator $O$ in the above expression can be
regarded to be the superconformal primary of spin $(\frac12 j,0)$, in which
case the result is summarized in \tablename~\ref{tab:WIObTO}. We can also
replace $O\to\Qb O$ of spin $(\frac12 j,\frac12)$ whose results are in \tablename~\ref{tab:WIQObTQbO}. Finally one could also consider $O \to QO$; the result is obtained by a simple rescaling of the coefficients in \tablename~\ref{tab:WIObTO} and a replacement $j\to j\pm 1$. For the reader's convenience we report here the relative normalizations for the operators in the $\cO$ multiplet as derived in \cite{Li:2014gpa}:
\eqn{
\frac{c_{(QO)^{(j+1,0)}}}{c_O} = 2\,\frac{j+2q}{(j+1)^2}\,,\quad \frac{c_{(QO)^{(j-1,0)}}}{c_O} = 2\,\frac{(j+1)(2q -j -2)}{j}\,,\quad \frac{c_{(\Qb O)}}{c_O}=4\,\qb\,.
}[twopointnorm]

\appendset{\begin{table}[!ht]
\centering
\begin{tabularx}{\textwidth}{XlllllX}
\hline
&$\cObTO{i}$ & Structure & $j>1$ & $j=1$ & $j=0$ &\\ \hline\hline
&$\cObTO1$
&\tableeqn{(\invJ132)^2\lsp (\invI13)^j}
&$\cObTO1$
&\tableeqn{\frac{i(2\Delta-3)}{3\pi^2}}
&\tableeqn{\frac{2\Delta}{3\pi^2}}
\\\hline
&$\cObTO2$
&\tableeqn{\invI12\lsp\invI23\lsp\invJ132\lsp(\invI13)^{j-1}}
&\tableeqn{-6\lsp\cObTO1 + \frac{4\llsp i^j(\Delta-j)}{\pi^2}}
&\tableeqn{\frac{2i}{\pi^2}}
&\hfill$\diagup$\hfill\lsp
\\\hline
&$\cObTO3$
&\tableeqn{(\invI12)^2\lsp(\invI23)^2\lsp(\invI13)^{j-2}}
&\tableeqn{6\lsp\cObTO1 - \frac{2\llsp i^j(2\Delta-3j)}{\pi^2}}
&\hfill$\diagup$\hfill\lsp
&\hfill$\diagup$\hfill\lsp
\\\hline
\end{tabularx}
\caption{Ward identities of the stress tensor for the correlator
$\langle\Ob T O\rangle$ when $O$ has spin $(\frac12 j,0)$. We have defined $\Delta = q+\qb$.}
\label{tab:WIObTO}
\end{table}}{wi}

\appendset{\begin{table}[!ht]
\centering
\small
\begin{tabularx}{\textwidth}{lllll}
\hline
$\cQObTQbO{i}$ & Structure & $j>1$ & $j=1$ & $j=0$ \\ \hline\hline
$\cQObTQbO1$
&\tableeqn{\invI23\lsp\invI32\lsp\invJ132\lsp\invJ231\lsp(\invI13)^{j-1}}
&\tableeqn{\cQObTQbO6}
&\tableeqn{\cQObTQbO6}
&\hfill$\diagup$\hfill\lsp
\\\hline
$\cQObTQbO2$
&\tableeqn{\invI31\lsp(\invJ132)^2\lsp(\invI13)^{j}}
&\tableeqn{\begin{aligned} &-\frac19\lsp(3\lsp\cQObTQbO5+2\lsp\cQObTQbO6+\cQObTQbO{10})\\
                          &\;-\frac16\lsp\cQObTQbO9-i^{j+1}\frac{2\Delta-j-2}{3\pi^2}
           \end{aligned}}
&\tableeqn{-\frac23\lsp\cQObTQbO6+\frac{2\Delta-5}{3\pi^2}}
&\tableeqn{-i\frac{2(\Delta-1)}{3\pi^2}}
\\\hline
$\cQObTQbO3$
&\tableeqn{\invI21\lsp\invI32\lsp\invJ132\lsp(\invI13)^{j}}
&\tableeqn{\begin{aligned} &-\frac19\lsp(3\lsp\cQObTQbO5+8\lsp\cQObTQbO6+\cQObTQbO{10})\\
                           &\;-\frac13\lsp\cQObTQbO9-\frac{2\llsp i^{j+1}(j-3)}{3\pi^2}
           \end{aligned}}
&\tableeqn{-\frac43\lsp\cQObTQbO6-\frac{2}{\pi^2}}
&\tableeqn{\frac{2i}{\pi^2}}
\\\hline
$\cQObTQbO4$
&\tableeqn{\invI12\lsp\invI32\lsp(\invI23)^2\lsp\invJ231(\invI13)^{j-2}}
&\tableeqn{\cQObTQbO{10}}
&\hfill$\diagup$\hfill\lsp
&\hfill$\diagup$\hfill\lsp
\\\hline
$\cQObTQbO5$
&\tableeqn{\invI12\lsp\invI31\lsp\invI23\lsp\invJ132\lsp(\invI13)^{j-1}}
&\tableeqn{\cQObTQbO5}
&\tableeqn{\frac43\lsp\cQObTQbO6+\frac{2}{\pi^2}}
&\hfill$\diagup$\hfill\lsp
\\\hline
$\cQObTQbO6$
&\tableeqn{\invI12\lsp\invI21\lsp\invJ132\lsp\invJ123\lsp(\invI13)^{j-1}}
&\tableeqn{\cQObTQbO6}
&\tableeqn{\cQObTQbO6}
&\hfill$\diagup$\hfill\lsp
\\\hline
$\cQObTQbO7$
&\tableeqn{\invI12\lsp\invI21\lsp\invI23\lsp\invI32\lsp(\invI13)^{j-1}}
&\tableeqn{\frac23\lsp(\cQObTQbO6-\cQObTQbO{10})}
&\tableeqn{\frac23\cQObTQbO6}
&\hfill$\diagup$\hfill\lsp
\\\hline
$\cQObTQbO8$
&\tableeqn{(\invI12)^2\lsp(\invI23)^2\lsp\invJ123\lsp\invJ231(\invI13)^{j-3}}
&\tableeqn{\begin{aligned} &-\frac43\lsp(2\lsp\cQObTQbO6+\cQObTQbO{10})\\
                           &\;+2\lsp(\cQObTQbO5+\cQObTQbO9)+\frac{4\llsp i^{j+1}\llsp j}{\pi^2}
           \end{aligned}}
&\hfill$\diagup$\hfill\lsp
&\hfill$\diagup$\hfill\lsp
\\\hline
$\cQObTQbO9$
&\tableeqn{(\invI12)^2\lsp(\invI23)^2\lsp\invI31\lsp(\invI13)^{j-2}}
&\tableeqn{\cQObTQbO9}
&\hfill$\diagup$\hfill\lsp
&\hfill$\diagup$\hfill\lsp
\\\hline
$\cQObTQbO{10}$
&\tableeqn{(\invI12)^2\lsp\invI21\lsp\invI23\lsp\invJ123\lsp(\invI13)^{j-2}}
&\tableeqn{\cQObTQbO{10}}
&\hfill$\diagup$\hfill\lsp
&\hfill$\diagup$\hfill\lsp
\\\hline
\end{tabularx}
\caption{Ward identities of the stress tensor for the correlator $\langle
\Ob' T O'\rangle$ when $O'$ has spin $(\frac12 j,\frac12)$, dimension $\Delta+\tfrac12$ and is assumed to be unit normalized. If $O' = \Qb O$ the terms not proportional to $\cQObTQbO{k}$ must be rescaled by $c_{(\Qb O)}$.}
\label{tab:WIQObTQbO}
\end{table}}{wi}

\paragraph{Supersymmetry current} For this Ward identity let us fix the
third operator to be $O$. We then have three choices:
$\mathsf{t}_{(Q\Ob)\overbar{S}O}$ and $\mathsf{t}_{(\Qb\Ob^\pm)SO}$, where we used $\Qb\Ob^\pm$ as a shorthand for $(\Qb\Ob)^{(j\pm1,0)}$. The topological operator obtained by integrating $S$ or $\overbar{S}$ over $\Sigma$ is precisely the supercharge $Q$ or $\Qb$ respectively. We thus readily obtain the following identities
\eqna{
  \frac{i}{2} \int_{\Sigma}\di \Omega(x_{23}) \,x_{23}{\!}^2
\,\partial_{\eta_2}\rmx_{23}\partial_{\etab_2}\,\mathsf{t}_{(Q\Ob)\overbar{S}O}(x_i;\eta_i,\etab_i)
&= \mathfrak{N}\,\etab_2\partial_{\etab_3}\,\mathsf{n}_{(Q\Ob)(\Qb O)}(x_{13},\eta_{1,3},\etab_{1,3}) \,,\\
\frac{i}{2} \int_{\Sigma}\di \Omega(x_{23}) \,x_{23}{\!}^2
\,\partial_{\eta_2}\rmx_{23}\partial_{\etab_2}\,\mathsf{t}_{(\Qb\Ob^+)SO}(x_i;\eta_i,\etab_i)
&= \mathfrak{N}\,\eta_2\partial_{\eta_3}\,\mathsf{n}_{(\Qb\Ob^+)(Q O^+)}(x_{13},\eta_{1,3},\etab_{1,3})\,,\\
\frac{i}{2} \int_{\Sigma}\di \Omega(x_{23}) \,x_{23}{\!}^2
\,\partial_{\eta_2}\rmx_{23}\partial_{\etab_2}\,\mathsf{t}_{(\Qb\Ob^-)SO}(x_i;\eta_i,\etab_i)
&= \mathfrak{N}\,\frac{j}{j+1}\eta_2\eta_3\,\mathsf{n}_{(\Qb\Ob^-)(Q O^-)}(x_{13},\eta_{1,3},\etab_{1,3})\,.
}[]
The two-point functions must be normalized according to \twopointnorm. All
the results are summarized in Tables~\ref{tab:WIQObSbO}, \ref{tab:WIQbObSOplus} and \ref{tab:WIQbObSOminus}.

\appendset{\begin{table}[!ht]
\centering
\begin{tabularx}{\textwidth}{lllll}
\hline
$\cQObSbO{i}$ & Structure & $j>1$ & $j=1$ & $j=0$ \\ \hline\hline
$\cQObSbO1$
& \tableeqn{\invJ231\lsp\invJ132\lsp\invI23\lsp(\invI13)^{j-1}}
& \tableeqn{\cQObSbO1}
& \tableeqn{\cQObSbO1}
&\hfill$\diagup$\hfill\lsp
\\\hline
$\cQObSbO2$
& \tableeqn{\invI21\lsp\invJ132\lsp(\invI13)^{j}}
& \tableeqn{-\frac{1}{2}(\cQObSbO1+\cQObSbO4) -\frac{1}{3}\cQObSbO3+ \frac{8\llsp i^j\llsp \qb}{3\pi^2}}
& \tableeqn{-\frac{1}{2}(\cQObSbO1+\cQObSbO4)+\frac{8i\qb}{3\pi^2}}
& \tableeqn{\frac{8\qb}{3\pi^2}}
\\\hline
$\cQObSbO3$
& \tableeqn{\invI12\lsp(\invI23)^2\lsp\invJ231\lsp(\invI13)^{j-2}}
& \tableeqn{\cQObSbO3}
&\hfill$\diagup$\hfill\lsp
&\hfill$\diagup$\hfill\lsp
\\\hline
$\cQObSbO4$
& \tableeqn{\invI12\lsp\invI21\lsp\invI23\lsp(\invI13)^{j-1}}
& \tableeqn{\cQObSbO4}
& \tableeqn{\cQObSbO4}
&\hfill$\diagup$\hfill\lsp
\\\hline
\end{tabularx}
\caption{Ward identities of the supersymmetry current for the correlator
$\langle (Q\Ob) \overbar{S} O\rangle$ when $Q\Ob$ has spin
$(\frac12,\frac12j)$.}
\label{tab:WIQObSbO}
\end{table}}{wi}

\appendset{\begin{table}[!ht]
\centering
\begin{tabularx}{\textwidth}{XlllllX}
\hline
&$\cQbObSOplus{i}$ & Structure & $j>1$ & $j=1$ & $j=0$ &\\ \hline\hline
&$\cQbObSOplus1$
& \tableeqn{\invJ132\lsp\invI12\lsp(\invI13)^j}
& \tableeqn{-\frac{2}{3}\lsp\cQbObSOplus2 - \frac{4\llsp i^j\llsp(2q+j)}{3\pi^2(j+1)}}
& \tableeqn{-\frac{2}{3}\lsp\cQbObSOplus2 - \frac{2i\llsp(2q+1)}{3\pi^2}}
& \tableeqn{-\frac{8 q}{3\pi^2}}
\\\hline
&$\cQbObSOplus2$
& \tableeqn{(\invI12)^2\lsp(\invI13)^{j-1}}
& \tableeqn{\cQbObSOplus2}
& \tableeqn{\cQbObSOplus2}
&\hfill$\diagup$\hfill\lsp
\\\hline
\end{tabularx}
\caption{Ward identities of the supersymmetry current for the correlator
$\langle (\Qb\Ob) S O\rangle$ when $\Qb \Ob$ has spin $(0,\frac12(j+1))$.}
\label{tab:WIQbObSOplus}
\end{table}}{wi}

\appendset{\begin{table}[!ht]
\centering
\begin{tabularx}{\textwidth}{XllllX}
\hline
&$\cQbObSOminus{i}$ & Structure & $j>1$ & $j=1$ &\\ \hline\hline
&$\cQbObSOminus1$
& \tableeqn{\invJ132\lsp\invK231\lsp(\invI13)^{j-1}}
& \tableeqn{-\frac{1}{3}\cQbObSOminus2 + \frac{4\llsp i^j\llsp(2q-j-2)}{3\pi^2}}
& \tableeqn{\frac{4 i\llsp(2q-3)}{3\pi^2}}
\\\hline
&$\cQbObSOminus2$
& \tableeqn{\invI12\lsp\invI23\lsp\invK231\lsp(\invI13)^{j-2}}
& \tableeqn{\cQbObSOminus2}
&\hfill$\diagup$\hfill\lsp
\\\hline
\end{tabularx}
\caption{Ward identities of the supersymmetry current for the correlator
$\langle (\Qb\Ob) S O\rangle$ when $\Qb \Ob$ has spin $(0,\frac12(j-1))$.}
\label{tab:WIQbObSOminus}
\end{table}}{wi}

\subsection{Shortening conditions}
\label{sec:shortening}
The possible shortening conditions on the superconformal multiplet $\cO$ have been classified in \cite{Cordova:2016emh}. In this section we will explore all of them. On the algebra generated by $Q$ we can have the shortening conditions
\begin{itemize}
\item[$L$:] Unconstrained action on $\cO$ (no null states). Unitarity bound
  $q \geq \frac12j+1$.
\item[$A_1$:] Null state $(QO)^{(j-1,\jb)}$, $j\geq 1$. Unitarity bound $q
  = \frac12j+1$.
\item[$A_2$:] Null state $(Q^2O)^{(0,\jb)}$, $j=0$. Unitarity bound $q =1$.
\item[$B$:] Null state $(QO)^{(1,\jb)}$, $j=0$. Unitarity bound $q =0$.
\end{itemize}
The same applies to the algebra generated by $\Qb$.
Therefore, a shortening condition on a superconformal multiplet can be
described by specifying a choice of $\cX_i = L,A_1,A_2,B$ for each of the
two subalgebras: $[\cX_L,\overbar{\cX}_R]$. For simplicity we will refer to
$[L,\overbar{B}]$ as \emph{chirality}. Furthermore the conditions $[\cX_L,\overbar{A}_1]$ are
absent because we are considering the case $\jb = 0$. Since $\cOb =
\cO^\dagger$, $\cOb$ will satisfy the conjugate shortening
$[\cX_R,\overbar{\cX}_L]$. However, after imposing reality, either one of the two conditions is sufficient.
\par
\paragraph{Shortening $\boldsymbol{B}$ or $\boldsymbol{\overbar{B}}$} For
the $\Qb O = 0$ case ($\qb = 0$) the prefactor of
\eqref{eq:threepfSuperspace} does not depend on $x_{\bar{3} 1}$, while for
the $QO = 0$ case ($q = j = 0$) the prefactor does not depend on
$x_{\bar{1}3}$. In both cases we can commute the superspace derivative and
obtain conditions on $t$ only. They read, respectively,
\eqn{
\etab_1\overbar{\cD}\,t(Z;\eta_i,\etab_i) = 0\,,\qquad
\eta_1\cD\,t(Z;\eta_i,\etab_i) = 0\,.
}[eq:chirality]
\paragraph{Shortening $\boldsymbol{A_1}$} Also in this case (when $q=j/2+1$) we can commute the differential operator with the prefactor due to
\eqn{
\partial_{\etab_1}\overbar{D}_1 \frac{(\eta_1\rmx_{3\bar{1}}\etab_1)^j}{{x_{\bar{1}3}}{\!}^{2j+2}} = 0\,,\qquad\mbox{for $x_{13}\neq 0$}\,,
}[]
and thus we readily obtain
\eqn{
\partial_{\eta_1}\lnsp\cD\,t(Z;\eta_i,\etab_i)=0\,.
}[eq:conservationO]
\paragraph{Shortening $\boldsymbol{A_2}$ or $\boldsymbol{\overbar{A}_2}$} In this case the commutation of the derivative and the prefactor is due to the identities
\eqn{
  {\overbar{D}_1}{\!}^2 \frac{1}{{x_{\bar{1}3}}{\!}^2} = {D_1}{\!}^2 \frac{1}{{x_{\bar{3}1}}{\!}^2}=0\,,\qquad\mbox{for $x_{13}\neq 0$}\,.
}[]
Thus for $Q^2O = 0$ ($q=1$ and $j=0$)  and for $\Qb^2 O = 0$ ($\qb = 1$) we get, respectively
\eqn{
\cD^2\,t(Z;\eta_i,\etab_i) = 0 \,,\qquad \overbar{\cD}^2\,t(Z;\eta_i,\etab_i) = 0\,.
}[eq:shortening]
\par In \tablename~\ref{tab:shortening} we summarize all the constraints arising from \eqref{eq:chirality}, \eqref{eq:conservationO} and \eqref{eq:shortening}. All shortening conditions can be easily obtained by combining them. \tablename~\ref{tab:final} instead shows how many independent coefficients are left in the superspace correlator as we choose different shortening conditions and impose all other constraints obtained before.
\begin{table}
\centering
\subtable[$j>1$]{
\begin{tabular}{l|lll|}
\multicolumn{1}{c}{} & \multicolumn{1}{c}{$\overbar{L}$} &
\multicolumn{1}{c}{$\overbar{A}_2$} & \multicolumn{1}{c}{$\bar{B}$} \\
\cline{2-4}
$L$ & 2 & 2 & 0\\
$A_1$ & 1 & 1 & \hfill$\diagup$\hfill\lsp \\
\cline{2-4}
\end{tabular}
}\qquad\quad
\subtable[$j=1$]{
\begin{tabular}{l|lll|}
\multicolumn{1}{c}{} & \multicolumn{1}{c}{$\overbar{L}$} & \multicolumn{1}{c}{$\overbar{A}_2$} & \multicolumn{1}{c}{$\overbar{B}$} \\
\cline{2-4}
$L$ & 1 & 1 & 0\\
$A_1$ & 1 & 1 & 0 \\
\cline{2-4}
\end{tabular}
}\qquad\quad
\subtable[$j=0$]{
\begin{tabular}{l|lll|}
\multicolumn{1}{c}{} & \multicolumn{1}{c}{$\overbar{L}$} & \multicolumn{1}{c}{$\overbar{A}_2$} & \multicolumn{1}{c}{$\overbar{B}$} \\
\cline{2-4}
$L$ & 0 & 0 & 0\\
$A_2$ & 0 & 0 & 0 \\
$B$ & 0 & 0 & $\boldsymbol0$ \\
\cline{2-4}
\end{tabular}
}
\caption{Number of independent coefficients $\cOJOi{k}$ of the superspace
correlator as different shortening conditions are chosen. The slash means
that there is no consistent three-point function. The boldface zero means
that the three-point function is identically zero. Other zeros imply that the three-point function is completely fixed in terms of $q$, $\qb$ and $j$. In all cases these numbers refer to \emph{real} degrees of freedom as the $\cOJOi{k}$ are either all real or all purely imaginary.}\label{tab:final}
\end{table}

\begin{table}
\centering
\begin{tabular}{cllll}
\hline
 & Constraints & Conditions \\ \hline\hline
$A_1$
&\tableeqn{\begin{aligned}
   &\cOJOF=(j-1)\cOJOC + \lifrac{j(j-1)}{j+1}(\cOJOE - 4\lsp\cOJOA)\,,\\
   &\cOJOG = -2\lsp\cOJOB + \cOJOC +j\lsp\cOJOD+\lifrac{2j}{j+1}(\cOJOE + (j-3)\lsp\cOJOA)\,,\\
   &\cOJOH = 4\lsp\cOJOB-\cOJOC +\lifrac{2j}{1+j}(4\lsp\cOJOA-\cOJOE)\,,\\
   &\cOJOJ = j\lsp\cOJOI = j\lsp\cOJOD + \lifrac{j}2(\cOJOC+\cOJOE)\,.
   \end{aligned}
  }
&$j\geq 1$, $q=j/2+1$
\\\hline
$A_2$
&\tableeqn{\begin{aligned}\cOJOI = \cOJOD + \lifrac12\lsp\cOJOE\,.\end{aligned}}
&$j=0$, $q=1$
\\\hline
$\overbar{A}_2$
&\tableeqn{\begin{aligned}
   &\cOJOI = -\lifrac12\lsp(\cOJOC+\cOJOE)-\cOJOD\,,\\
   &\cOJOJ = -\lifrac12\lsp(\cOJOF+\cOJOH)-\cOJOG\,.
   \end{aligned}}
&$\qb=1$
\\\hline
$B$
&\tableeqn{\begin{aligned}\cOJOD = -2\lsp \cOJOA\,,\quad \cOJOE = 4\lsp\cOJOA\,,\quad \cOJOI=0\,.\end{aligned}}
&$j=0$, $q=0$
\\\hline
$\overbar{B}$
& \tableeqn{\begin{aligned}
   &\cOJOD = 2\lsp\cOJOA\,,\quad \cOJOE = -4\lsp\cOJOA\,,\\
   &\cOJOG = 2\lsp\cOJOB\,,\quad \cOJOH = -4\lsp\cOJOB\,,\\
   &\cOJOC=\cOJOF=\cOJOI=\cOJOJ=0\,.
   \end{aligned}
  }
&$\qb=0$
\\\hline
\end{tabular}
\caption{Constraints on the coefficients $\cOJOi{k}$ following from the various shortening conditions on the multiplet $\cO$ (here $\jb=0$ is implicit). Case $A_1$ for $j=1$ and cases $\overbar{A}_2$ and $\overbar{B}$ for $j=0,1$ can be obtained by setting to zero the absent coefficients ($\cOJOF$ for $j=1$ and $\cOJOi{2,3,6,7,8,10}$ for $j=0$).}
\label{tab:shortening}
\end{table}

\section{Expansion of the superspace correlator}
\label{sec:expansion}

In order to apply the various constraints originating from the ANEC to our
three-point function in superspace we need to express its components in a
basis of nonsupersymmetric three-point functions. This will be achieved by
Taylor expanding in the Grassmann coordinates $\theta_i,\thetab_i$. We
relied on a \emph{Mathematica} package\footnote{Which can be made available
upon request.} to perform the spinor algebra involved in this computation.
Due to the Schouten identities mentioned above, it is hard to determine whether two quantities are equal. Therefore we check for equality by replacing the various quantities that appear with random numerical values.\footnote{After sufficiently many replacements, this is equivalent to picking a basis at random and checking for equality for every vector in it. The fact that we replace numerical values to Grassmann coordinates is not an issue if one orders the factors in a canonical way before applying the replacement. Moreover there are no precision issues because we use exact rational numbers.}
\par
Every order that contains at least a $\theta$ and a $\thetab$ at the same
point will mix with conformal descendants due to $\{Q_\alpha,\Qb_\alphad\}
= 2\llsp\mathrm{P}_{\alpha\alphad}$. The results of \cite{Li:2014gpa} can be used to subtract these contributions.
We will only perform this expansion to first order in $\theta_i,\thetab_i$ and
not for all possible combinations but only the ones of interest. We also
performed the expansion to all orders in $\theta_2,\thetab_2$ and to all
orders in $\theta_1,\thetab_1$ to make some consistency
checks,\footnote{Namely we observed that the order
$\theta_2^2\thetab_2{\!\!\hspace{-0.6pt}}^2$ consists only of descendants when the conservation condition \eqref{eq:consCond} is applied, consistently with the operator content of $\cJ$. In addition we verified that applying the shortening differential operators in \sectionname~\ref{sec:shortening} on the expanded correlator yields the same constraints.} but we will not present these results here. For nonsupersymmetric three-point functions we will remain consistent with the conventions introduced in \sectionname~\ref{sec:WI}.

\subsection{Lowest order}

At this order we simply have $J$. Consistently with the previous sections
we denote the three-point function coefficients by
\eqn{
\mathsf{t}_{\Ob J O} \;\longrightarrow \;\cObJO{k}\,.
}[]
The results, without assuming the reality condition and conservation, are shown in \tablename~\ref{tab:matchingObJO}.

\appendset{\begin{table}[!ht]
\centering
\begin{tabularx}{\textwidth}{XllllX}
\hline
&$\cObJO{i}$ & Structure & $j>0$ & $j=0$ &\\ \hline\hline
&$\cObJO{1}$ &
\tableeqn{\invJ132\lsp(\invI13)^j} &
\tableeqn{i(\cOJOA+\cOJOB)} &
\tableeqn{i\lsp\cOJOA}
\\\hline
&$\cObJO{2}$ &
\tableeqn{\invI23\lsp(\invI13)^{j-1}} &
\tableeqn{-i\lsp\cOJOB} &
\hfill\tableeqn{\diagup}\hfill\lsp\\
\hline
\end{tabularx}
\caption{Expansion of the supersymmetric correlator in the component
$\langle\Ob J O\rangle$ when $O$ has spin $(\frac12 j,0)$.}
\label{tab:matchingObJO}
\end{table}}{nonsusy}

\subsection{Three-point function \texorpdfstring{$\langle \Ob T O\rangle$}{<Ob T O>}}

At order $\theta_2\thetab_2$ we have the
stress-energy tensor. Consistently with the
previous sections we denote the three-point function coefficients by
\eqn{
\mathsf{t}_{\Ob T O} \;\longrightarrow \; \cObTO{k}\,.
}[]
The results are shown in \tablename~\ref{tab:matchingObTO}. The conservation of $\cJ$ and the reality condition are not assumed there. In principle the expansion also contains superdescendants of $\cJ$ of spin $(0,0),\,(0,1)$ and $(1,0)$. We checked that those contributions vanish after imposing conservation and we will not report those results here.

\appendset{\begin{table}[!ht]
\centering
\begin{tabularx}{\textwidth}{XlllllX}
\hline
&$\cObTO{i}$ & Structure & $j>1$ & $j=1$ & $j=0$ & \\ \hline\hline
&$\cObTO1$
&\tableeqn{(\invJ132)^2\lsp (\invI13)^j}
&\tableeqn{-\frac14(\cOJOE+\cOJOH)}
&\tableeqn{-\frac14(\cOJOE+\cOJOH)}
&\tableeqn{-\frac14\lsp\cOJOE}
\\\hline
&$\cObTO2$
&\tableeqn{\invI12\lsp\invI23\lsp\invJ132\lsp(\invI13)^{j-1}}
&\tableeqn{\frac14(\cOJOF+\cOJOH)}
&\tableeqn{\frac14\lsp\cOJOH}
&\hfill$\diagup$\hfill\lsp
\\\hline
&$\cObTO3$
&\tableeqn{(\invI12)^2\lsp(\invI23)^2\lsp(\invI13)^{j-2}}
&\tableeqn{-\frac14\lsp\cOJOF}
&\hfill$\diagup$\hfill\lsp
&\hfill$\diagup$\hfill\lsp
\\\hline
\end{tabularx}
\caption{Expansion of the supersymmetric correlator in the component
$\langle\Ob T O\rangle$ when $O$ has spin $(\frac12 j,0)$.}
\label{tab:matchingObTO}
\end{table}}{nonsusy}

\subsection{Three-point functions \texorpdfstring{$\langle (\Qb\Ob) S
O\rangle$}{<QbOb S O>} and \texorpdfstring{$\langle (Q\Ob) \overbar{S} O\rangle$}{<QOb Sb O>}}

At order $\theta_1\thetab_2$, $\thetab_1\theta_2$ we have the supersymmetry current with the first superdescendant of $\cOb$. The naming of the coefficients is
\eqn{
\mathsf{t}_{(\Qb\Ob^+) S O} \;\longrightarrow \; \cQbObSOplus{k}\,,\qquad
\mathsf{t}_{(\Qb\Ob^-) S O} \;\longrightarrow \; \cQbObSOminus{k}\,,\qquad
\mathsf{t}_{(Q\Ob) \overbar{S} O} \;\longrightarrow \; \cQObSbO{k}\,.\qquad
}
As before $\Qb\Ob^\pm$ stands for $(\Qb\Ob)^{(0,j\pm1)}$. Also in these
cases the results are presented without conservation and reality
applied---they can be found in Tables~\ref{tab:matchingQbObSOp}, \ref{tab:matchingQbObSOm} and \ref{tab:matchingQObSbO}. There are also contributions from superdescendants of spin $(0,\frac12)$ or $(\frac12,0)$. As in the previous subsection we have verified that they vanish after conservation is imposed and we will not report those results.

\appendset{\begin{table}[!ht]
\centering
\begin{tabularx}{\textwidth}{XllllX}
\hline
&$\cQbObSOplus{i}$ & Structure & $j>0$ & $j=0$ &\\ \hline\hline
&$\cQbObSOplus1$
& \tableeqn{\invJ132\lsp\invI12\lsp(\invI13)^j}
& \tableeqn{-\frac{1}{2(1+j)}\left(4\lsp\cOJOA+4\lsp\cOJOB +\cOJOC-\cOJOE+\cOJOF-\cOJOH\right)}
& \tableeqn{-2\lsp\cOJOA-\frac12\lsp\cOJOE}
\\\hline
&$\cQbObSOplus2$
& \tableeqn{(\invI12)^2\lsp(\invI13)^{j-1}}
& \tableeqn{\frac{1}{2(1+j)}\left(4\lsp\cOJOB + \cOJOC+\cOJOF-\cOJOH\right)}
&\hfill$\diagup$\hfill\lsp
\\\hline
\end{tabularx}
\caption{Expansion of the supersymmetric correlator in the component
$\langle (\Qb\Ob) S O\rangle$ when $\Qb \Ob$ has spin $(0,\frac12(j+1))$. The result for $j=1$ is obtained by setting $\cOJOF = 0$.}
\label{tab:matchingQbObSOp}
\end{table}}{nonsusy}

\appendset{\begin{table}[!ht]
\centering
\begin{tabularx}{\textwidth}{XllllX}
\hline
&$\cQbObSOminus{i}$ & Structure & $j>1$ & $j=1$ &\\ \hline\hline
&$\cQbObSOminus1$
& \tableeqn{\invJ132\lsp\invK231\lsp(\invI13)^{j-1}}
& \tableeqn{2(\cOJOA+\cOJOB)-\frac12(\cOJOE+\cOJOH)-\frac{1}{2j}(\cOJOC+\cOJOF)}
& \tableeqn{
}
\\\hline
&$\cQbObSOminus2$
& \tableeqn{\invI12\lsp\invI23\lsp\invK231\lsp(\invI13)^{j-2}}
& \tableeqn{\frac{1}{j}\cOJOF-\frac{j-1}{2j}\left(4\lsp\cOJOB+\cOJOC-\cOJOH\right)}
&\hfill$\diagup$\hfill\lsp
\\\hline
\end{tabularx}
\caption{Expansion of the supersymmetric correlator in the component
$\langle (\Qb\Ob) S O\rangle$ when $\Qb \Ob$ has spin $(0,\frac12(j-1))$. The unbarred entry in the $j=1$ column is obtained by setting $\cOJOF = 0$.}
\label{tab:matchingQbObSOm}
\end{table}}{nonsusy}

\appendset{\begin{table}[!ht]
\centering
\begin{tabularx}{\textwidth}{XlllllX}
\hline
&$\cQObSbO{i}$ & Structure & $j>1$ & $j=1$ & $j=0$ &\\ \hline\hline
&$\cQObSbO1$
& \tableeqn{\invJ231\lsp\invJ132\lsp\invI23\lsp(\invI13)^{j-1}}
& \tableeqn{\cOJOB-\frac12\lsp\cOJOG}
&
&\hfill$\diagup$\hfill\lsp
\\\hline
&$\cQObSbO2$
& \tableeqn{\invI21\lsp\invJ132\lsp(\invI13)^{j}}
& \tableeqn{-2(\cOJOA+\cOJOB)-\frac12(\cOJOE+\cOJOH)}
&
& \tableeqn{-2\lsp\cOJOA-\frac12\lsp\cOJOE}
\\\hline
&$\cQObSbO3$
& \tableeqn{\invI12\lsp(\invI23)^2\lsp\invJ231\lsp(\invI13)^{j-2}}
& \tableeqn{-\frac12\lsp\cOJOF}
&\hfill$\diagup$\hfill\lsp
&\hfill$\diagup$\hfill\lsp
\\\hline
&$\cQObSbO4$
& \tableeqn{\invI12\lsp\invI21\lsp\invI23\lsp(\invI13)^{j-1}}
& \tableeqn{\cOJOB+\frac12(\cOJOF+\cOJOG+\cOJOH)}
&
&\hfill$\diagup$\hfill\lsp
\\\hline
\end{tabularx}
\caption{Expansion of the supersymmetric correlator in the component
$\langle (Q\Ob) \overbar{S} O\rangle$ when $Q\Ob$ has spin
$(\frac12,\frac12 j)$. The unbarred entries in the $j=1$ column can be obtained by setting the $\cOJOF=0$.}
\label{tab:matchingQObSbO}
\end{table}}{nonsusy}

\subsection{Three-point functions \texorpdfstring{$\langle (\Qb\Ob) J (QO)\rangle$}{<QbOb J QO>} and \texorpdfstring{$\langle (Q\Ob) J (\Qb O)\rangle$}{<QOb J QbO>}}

At order $\theta_1\thetab_3$, $\thetab_1\theta_3$ we extract the descendants $QO,\Qb O$ and their conjugates. We need this mainly as a preliminary result for the computation of the next subsection. We named
\eqna{
&\mathsf{t}_{(\Qb\Ob^+) J (QO^+)} \;\longrightarrow \; \cQbObJQOpp{k}\,,\qquad
\mathsf{t}_{(\Qb\Ob^+) J (QO^-)} \;\longrightarrow \; \cQbObJQOpm{k}\,,\qquad\\
&\mathsf{t}_{(\Qb\Ob^-) J (QO^+)} \;\longrightarrow \; \cQbObJQOmp{k}\,,\qquad
\mathsf{t}_{(\Qb\Ob^-) J (QO^-)} \;\longrightarrow \; \cQbObJQOmm{k}\,,
}[]
where $(Q O^\pm)$ stands for $(Q O)^{(j\pm1,0)}$, and
\eqn{
\mathsf{t}_{(Q\Ob) J (\Qb O)} \;\longrightarrow \; \cQObJQbO{k}\,.
}[]
In order to make the computation more manageable, this time we applied
conservation and reality from the start. The results are in Tables~\ref{tab:matchingQbObJQOpp}, \ref{tab:matchingQbObJQOpm}, \ref{tab:matchingQbObJQOmp}, \ref{tab:matchingQbObJQOmm} and \ref{tab:matchingQObJQbO}.

\appendset{\begin{table}[!ht]
\centering
\begin{tabularx}{\textwidth}{XlllX}
\hline
&$\cQbObJQOpp{i}$ & Structure & $j\geq 0$ &\\ \hline\hline
&$\cQbObJQOpp1$
& \tableeqn{\invJ132\lsp(\invI13)^{j+1}}
& \tableeqn{-\frac{2(2q+j-1)}{(j+1)^2}(\cOJOA+\cOJOB) - \frac{1}{(j+1)^2}(\cOJOD+\cOJOE+\cOJOG+\cOJOH)}
\\\hline
&$\cQbObJQOpp2$
& \tableeqn{\invI12\lsp\invI23\lsp(\invI13)^j}
& \tableeqn{\frac{1}{(j+1)^2}\left(2\lsp\cOJOA+2(2q+j-1)\lsp\cOJOB + \cOJOD+\cOJOG+\cOJOH\right)}
\\\hline
\end{tabularx}
\caption{Expansion of the supersymmetric correlator in the component
  $\langle (\Qb\Ob) J (QO)\rangle$ when $\Qb \Ob$ has spin $(0,\frac12(j+1))$ and $QO$
has spin $(\frac12(j+1),0)$. The result for $j=0,1$ can be obtained by setting the absent coefficients to zero (see caption of \tablename~\ref{tab:shortening}).}
\label{tab:matchingQbObJQOpp}
\end{table}}{nonsusy}

\appendset{\begin{table}[!ht]
\centering
\begin{tabularx}{\textwidth}{llllX}
\hline
&$\cQbObJQOpm{i}$ & Structure & $j\geq 1$ \\ \hline\hline
&$\cQbObJQOpm1$
& \tableeqn{\invI12\lsp\invKb123\lsp(\invI13)^{j-1}}
& \tableeqn{\frac{1}{j(j+1)}\left(4(q-1)\lsp\cOJOB+\cOJOH\right) - \frac{1}{j+1}(2\lsp\cOJOA+\cOJOD+\cOJOG)-\frac{1}{j}(\cOJOC+\cOJOF)} &
\\\hline
\end{tabularx}
\caption{Expansion of the supersymmetric correlator in the component
  $\langle (\Qb\Ob) J (QO)\rangle$ when $\Qb \Ob$ has spin $(0,\frac12(j+1))$ and $QO$
has spin $(\frac12(j-1),0)$. The result for $j=1$ can be obtained by setting $\cOJOF$ to zero.}
\label{tab:matchingQbObJQOpm}
\end{table}}{nonsusy}

\appendset{\begin{table}[!ht]
\centering
\begin{tabularx}{\textwidth}{XlllX}
\hline
&$\cQbObJQOmp{i}$ & Structure & $j\geq1$\\ \hline\hline
&$\cQbObJQOmp1$
& \tableeqn{\invI23\lsp\invK231\lsp(\invI13)^{j-1}}
& \tableeqn{\frac{1}{j(j+1)}\left(2(2q-j-3)\lsp\cOJOB+\cOJOH+\cOJOG\right) - \frac{1}{j+1}(2\lsp\cOJOA+\cOJOD)} &
\\\hline
\end{tabularx}
\caption{Expansion of the supersymmetric correlator in the component
  $\langle (\Qb\Ob) J (QO)\rangle$ when $\Qb \Ob$ has spin $(0,\frac12(j-1))$ and $QO$
has spin $(\frac12(j+1),0)$.}
\label{tab:matchingQbObJQOmp}
\end{table}}{nonsusy}

\appendset{\begin{table}[!ht]
\centering
\begin{tabularx}{\textwidth}{llllX}
\hline
&\hspace{-3ex}$\cQbObJQOmm{i}$ & Structure & $j\geq 1$ \\ \hline\hline
&\hspace{-3ex}$\cQbObJQOmm1$
& \tableeqn{\invJ132\lsp(\invI13)^{j-1}}
& \tableeqn{-\frac{2 \,\Xi_1}{j}\lsp \cOJOA-\frac{2\,\Xi_2}{j^2}\lsp\cOJOB-\frac{1}{j^2}\lsp\cOJOG+\frac1j\lsp\cOJOD+\frac{j+1}{j^2}(\cOJOC+\cOJOF+j\lsp\cOJOE)+\frac{j^2+j-1}{j^2}\lsp\cOJOH} &
\\\hline
&\hspace{-3ex}$\cQbObJQOmm2$
& \tableeqn{\invI12\lsp\invI23\lsp(\invI13)^{j-2}}\hspace{-1.3em}
& \tableeqn{\begin{aligned}
            &\frac{2(j-1)\,\Xi_3}{j^2}\lsp\cOJOB + \frac{j-1}{j}(2\lsp\cOJOA +\cOJOD) + \frac{j^2-1}{j^2}\lsp\cOJOC - \frac{2(j+1)}{j^2}\lsp \cOJOF-\frac{j-1}{j^2}\lsp\cOJOG\\
            &\;-\frac{(j-1)(j+2)}{j^2}\lsp\cOJOH
            \end{aligned}} &
\\\hline
\end{tabularx}
\caption{Expansion of the supersymmetric correlator in the component
  $\langle (\Qb\Ob) J (QO)\rangle$ when $\Qb \Ob$ has spin $(0,\frac12(j-1))$ and
  $QO$ has spin $(\frac12(j-1),0)$. The result for $j=1$ can be obtained by setting $\cOJOF$ to zero and removing the last row. Furthermore we defined
}
\vspace{-\baselineskip}
\eqna{
\Xi_1 &= j^2-2 j q+5 j-2 q+3\,,\\
\Xi_2 &= j^3-2 j^2 q+5 j^2-2 j q+3 j+2 q-3\,,\\
\Xi_3 &= j^2-2 j q+6 j-4 q+7\,.
}
\label{tab:matchingQbObJQOmm}
\end{table}}{nonsusy}

\appendset{\begin{table}[!ht]
\centering
\begin{tabularx}{\textwidth}{llllllX}
\hline
&\hspace{-1ex}$\cQObJQbO{i}$ & Structure & $j>1$ & $j=1$ & $j=0$ &\\ \hline\hline
&\hspace{-1ex}$\cQObJQbO1$
& \tableeqn{\invJ231\lsp\invI23\lsp\invI32\lsp(\invI13)^{j-1}}
& \tableeqn{\cOJOG-2\lsp\cOJOB}
&
&\hfill$\diagup$\hfill\lsp
\\\hline
&\hspace{-1ex}$\cQObJQbO2$
& \tableeqn{\invJ132\lsp\invI31\lsp(\invI13)^j}
& \tableeqn{\begin{aligned}
			&2(2\qb-1)(\cOJOA+\cOJOB)\\&
			-\cOJOD-\cOJOE-\cOJOG-\cOJOH
			\end{aligned}}
&
& \tableeqn{2(2\qb-1)\cOJOA-\cOJOD-\cOJOE}
\\\hline
&\hspace{-1ex}$\cQObJQbO3$
& \tableeqn{\invI21\lsp\invI32\lsp(\invI13)^j}
& \tableeqn{2(\cOJOA+\cOJOB)-\cOJOD-\cOJOG}
&
& \tableeqn{2\lsp\cOJOA-\cOJOD}
\\\hline
&\hspace{-1ex}$\cQObJQbO4$
& \tableeqn{\invJ231\lsp\invJ123\lsp\invI12\lsp\invI23\lsp(\invI13)^{j-2}}
& \tableeqn{\cOJOF}
& \hfill$\diagup$\hfill\lsp
&\hfill$\diagup$\hfill\lsp
\\\hline
&\hspace{-1ex}$\cQObJQbO5$
& \tableeqn{\invI12\lsp\invI23\lsp\invI31\lsp(\invI13)^{j-1}}
& \tableeqn{-2(2\qb-1)\cOJOB-\cOJOC+\cOJOG+\cOJOH}
&
&\hfill$\diagup$\hfill\lsp
\\\hline
&\hspace{-1ex}$\cQObJQbO6$
& \tableeqn{\invJ123\lsp\invI12\lsp\invI21(\invI13)^{j-1}}
& \tableeqn{-\cOJOC-\cOJOF}
&
&\hfill$\diagup$\hfill\lsp
\\\hline
\end{tabularx}
\caption{Expansion of the supersymmetric correlator in the component
$\langle (Q\Ob) J (\Qb O)\rangle$ when $Q\Ob$ has spin $(\frac12,\frac12 j)$. The unbarred entries in the $j=1$ column can be obtained by setting $\cOJOF=0$.}
\label{tab:matchingQObJQbO}
\end{table}}{nonsusy}

\subsection{Three-point functions \texorpdfstring{$\langle (\Qb\Ob) T (QO)\rangle$}{<QbOb T QO>} and \texorpdfstring{$\langle (Q\Ob) T (\Qb O)\rangle$}{<QOb T QbO>}}

At order $\theta_1\theta_2\thetab_2\thetab_3$,
$\thetab_1\theta_2\thetab_2\theta_3$ we extract the descendants $QO,\Qb O$
and their conjugates coupled with the stress tensor.  These terms are needed in order to impose the ANEC on superconformal descendants inside $\cO$. We named
\eqna{
&\mathsf{t}_{(\Qb\Ob^+) T (QO^+)} \;\longrightarrow \; \cQbObTQOpp{k}\,,\qquad
\mathsf{t}_{(\Qb\Ob^+) T (QO^-)} \;\longrightarrow \; \cQbObTQOpm{k}\,,\qquad\\
&\mathsf{t}_{(\Qb\Ob^-) T (QO^+)} \;\longrightarrow \; \cQbObTQOmp{k}\,,\qquad
\mathsf{t}_{(\Qb\Ob^-) T (QO^-)} \;\longrightarrow \; \cQbObTQOmm{k}\,,
}[]
\eqn{
\mathsf{t}_{(Q\Ob) T (\Qb O)} \;\longrightarrow \; \cQObTQbO{k}\,.
}[]
Also this time we applied conservation and reality from the start. The
results are in Tables~\ref{tab:matchingQbObTQOpp}, \ref{tab:matchingQbObTQOpm}, \ref{tab:matchingQbObTQOmp}, \ref{tab:matchingQbObTQOmm} and \ref{tab:matchingQObTQbO}.

\appendset{\begin{table}[!ht]
\centering
\begin{tabularx}{\textwidth}{lllllX}
\hline
&\hspace{-2ex}$\cQbObTQOpp{i}$ & Structure & $j>0$  & $j=0$ &\\ \hline\hline
&\hspace{-2ex}$\cQbObTQOpp1$
& \tableeqn{(\invJ132)^2\lsp(\invI13)^{j+1}}
& \tableeqn{\begin{aligned}
			&-\frac{i}{2(j+1)^2}\big(4\lsp\cOJOA+(2q+j)(\cOJOC+\cOJOF)\\
			&\;-(2q+j-2)(4\lsp\cOJOB+2\lsp\cOJOD)\big)
			\end{aligned}}
& \tableeqn{-2i\lsp\cOJOA+i(2q-1)\cOJOD}
\\\hline
&\hspace{-2ex}$\cQbObTQOpp2$
& \tableeqn{\invJ132\lsp\invI12\lsp\invI23\lsp(\invI13)^j}
& \tableeqn{\begin{aligned}
			&\frac{i	}{(j+1)^2}\big(6\lsp\cOJOA+3\lsp\cOJOD-2(2q+j-10)\cOJOB\\
			&\;+(2q+j-1)\cOJOC+(2q+j)\cOJOF\big)
			\end{aligned}}
& \tableeqn{3i(2\lsp\cOJOA+\cOJOD)}
\\\hline
&\hspace{-2ex}$\cQbObTQOpp3$
& \tableeqn{(\invI12)^2\lsp(\invI23)^2\lsp(\invI13)^{j-1}}
& \tableeqn{-\frac{i}{2(j+1)^2}\big(32\lsp\cOJOB-4\lsp\cOJOC+(2q+j)\cOJOF\big)}
&\hfill$\diagup$\hfill\lsp
\\\hline
\end{tabularx}
\caption{Expansion of the supersymmetric correlator in the component
  $\langle (\Qb\Ob) T (QO)\rangle$ when $\Qb \Ob$ has spin $(0,\frac12(j+1))$ and $QO$
has spin $(\frac12(j+1),0)$. The result for $j=1$ can be obtained by setting $\cOJOF=0$.}
\label{tab:matchingQbObTQOpp}
\end{table}}{nonsusy}

\appendset{\begin{table}[!ht]
\centering
\begin{tabularx}{\textwidth}{XllllX}
\hline
&$\cQbObTQOpm{i}$ & Structure & $j>1$ & $j=1$ &\\ \hline\hline
&$\cQbObTQOpm1$
& \tableeqn{\invKb123\lsp\invJ132\lsp\invI12\lsp(\invI13)^{j-1}}
& \tableeqn{\begin{aligned}
			&-\frac{3i}{j+1}(2\lsp\cOJOA+\lsp\cOJOD)+\frac{i}{j(j+1)}\big((2q+j-1)\lsp\cOJOC\\
			&\;-(2q+j)\lsp\cOJOF+2(2q+7j-4)\lsp\cOJOB\big)
			\end{aligned}}
& \tableeqn{}
\\\hline
&$\cQbObTQOpm2$
& \tableeqn{\invKb123\lsp\invI23\lsp(\invI12)^2\lsp(\invI13)^{j-2}}
& \tableeqn{\frac{i}{j(j+1)}\big(2(j-1)(8\lsp\cOJOB-\cOJOC)-(2q+j)\lsp\cOJOF\big)}
&\hfill$\diagup$\hfill\lsp
\\\hline
\end{tabularx}
\caption{Expansion of the supersymmetric correlator in the component
  $\langle (\Qb\Ob) T (QO)\rangle$ when $\Qb \Ob$ has spin $(0,\frac12(j+1))$ and $QO$
has spin $(\frac12(j-1),0)$. The unbarred entry in the $j=1$ column can be obtained by setting $\cOJOF=0$.}
\label{tab:matchingQbObTQOpm}
\end{table}}{nonsusy}

\appendset{\begin{table}[!ht]
\centering
\begin{tabularx}{\textwidth}{XllllX}
\hline
&$\cQbObTQOmp{i}$ & Structure & $j>1$ & $j=1$ &\\ \hline\hline
&$\cQbObTQOmp1$
& \tableeqn{\invK231\lsp\invJ132\lsp\invI23\lsp(\invI13)^{j-1}}
& \tableeqn{\begin{aligned}
			&-\frac{3i}{j+1}(2\lsp\cOJOA+\lsp\cOJOD)+\frac{i}{j(j+1)}\big((2q+j-1)\lsp\cOJOC\\
			&\;-(2q+j)\lsp\cOJOF-2(2q+7j-4)\lsp\cOJOB\big)
			\end{aligned}}
& \tableeqn{}
\\\hline
&$\cQbObTQOmp2$
& \tableeqn{\invK231\lsp\invI12\lsp(\invI23)^2\lsp(\invI13)^{j-2}}
& \tableeqn{\frac{i}{j(j+1)}\big(2(j-1)(8\lsp\cOJOB-\cOJOC)-(2q+j)\lsp\cOJOF\big)}
&\hfill$\diagup$\hfill\lsp
\\\hline
\end{tabularx}
\caption{Expansion of the supersymmetric correlator in the component
$\langle (\Qb\Ob) T (QO)\rangle$ when $\Qb \Ob$ has spin $(0,\frac12(j-1))$ and $QO$
has spin $(\frac12(j+1),0)$. The unbarred entry in the $j=1$ column can be obtained by setting $\cOJOF=0$. Note that this Table is identical to \tablename~\ref{tab:matchingQbObTQOpm}.}
\label{tab:matchingQbObTQOmp}
\end{table}}{nonsusy}

\appendset{\begin{table}[!ht]
\centering
\begin{tabularx}{\textwidth}{llllllX}
\hline
&\hspace{-2ex}$\cQbObTQOmm{i}$ & Structure & $j>2$ & $j=2$ & $j=1$ &\\ \hline\hline
&\hspace{-2ex}$\cQbObTQOmm1$
& \tableeqn{(\invJ132)^2\lsp(\invI13)^{j-1}}
& \hspace{-2ex}\tableeqn{\begin{aligned}
			&-\frac{2i(2j-1)}{j}\lsp\cOJOA +\frac{2 i \lsp \Xi_4}{j^2}\lsp\cOJOB - \frac{i\lsp \Xi_5}{2j^2}\lsp\cOJOC\\
			&\;+\frac{i\lsp\Xi_6}{j}\lsp\cOJOD - \frac{i(j-1)(\Xi_6 - 2q + j - 1)}{2j^2}\lsp\cOJOF
			\end{aligned}}
&
& \tableeqn{}
\\\hline
&\hspace{-2ex}$\cQbObTQOmm2$
& \tableeqn{\invJ132\lsp\invI12\lsp\invI23\lsp(\invI13)^{j-2}}
& \hspace{-2ex}\tableeqn{\begin{aligned}
			&\frac{6i(j-1)}{j}\lsp\cOJOA-\frac{2i(j-1)\lsp\Xi_7}{j^2}\lsp\cOJOB + \frac{3i(j-1)}{j}\lsp\cOJOD\\
			&\;\,+ \frac{i(j-1)(\Xi_7+9j-12)}{j^2}\lsp\cOJOC  + \frac{i \lsp\Xi_8}{j^2}\lsp\cOJOF
			\end{aligned}}
&
&\hfill$\diagup$\hfill\lsp
\\\hline
&\hspace{-2ex}$\cQbObTQOmm3$
& \tableeqn{(\invI12)^2\lsp(\invI23)^2\lsp(\invI13)^{j-3}}
& \hspace{-2ex}\tableeqn{\begin{aligned}
			&\frac{2i(j-1)(j-2)}{j^2}(\cOJOC-8\lsp\cOJOB)\\
			&\;- \frac{i(j-2)(j^2-2 j q+j-6 q+2)}{2j^2}\lsp\cOJOF
			\end{aligned}}
&\hfill$\diagup$\hfill\lsp
&\hfill$\diagup$\hfill\lsp
\\\hline
\end{tabularx}
\caption{Expansion of the supersymmetric correlator in the component
  $\langle (\Qb\Ob) T (QO)\rangle$ when $\Qb \Ob$ has spin $(0,\frac12(j-1))$ and $QO$
has spin $(\frac12(j-1),0)$. The unbarred entries in the $j=2$ column are identical and the ones in the $j=1$ column are obtained by setting $\cOJOF=0$. We further defined:}
\vspace{-\baselineskip}
\eqna{
\Xi_4 &= j^3-2 j^2 q-j^2-2 j q+5 j+2 q-4\,,\\
\Xi_5 &= j^3-2 j^2 q+j^2-2 j q+4 q-4\,,\\
\Xi_6 &= j^2-2 j q+j-2 q+3\,,\\
\Xi_7 &= j^2-2 j q-8 j-4 q+18\,, \\
\Xi_8 &= j^3-2 j^2 q-2 j q+8 q-3\,.
}
\label{tab:matchingQbObTQOmm}
\end{table}}{nonsusy}

\appendset{\begin{table}[!ht]
\centering
\begin{tabularx}{\textwidth}{llllllX}
\hline
&$\cQObTQbO{i}$ & Structure & $j>1$ & $j=1$ & $j=0$ &\\ \hline\hline
&$\cQObTQbO1$
&\tableeqn{\invI23\lsp\invI32\lsp\invJ132\lsp\invJ231\lsp(\invI13)^{j-1}}
&\tableeqn{\frac{3i}2(\cOJOC+\cOJOF)}
&\tableeqn{
}
&\hfill$\diagup$\hfill\lsp
\\\hline
&$\cQObTQbO2$
&\tableeqn{\invI31\lsp(\invJ132)^2\lsp(\invI13)^{j}}
&\tableeqn{\begin{aligned}
		   &-2i(\cOJOA+2\qb\lsp \cOJOB)+i(\qb-1)(\cOJOC+\cOJOF)\\
		   &\;-i(2\qb-1)\cOJOD
		   \end{aligned}}
&\tableeqn{
		   }
&\tableeqn{
}
\\\hline
&$\cQObTQbO3$
&\tableeqn{\invI21\lsp\invI32\lsp\invJ132\lsp(\invI13)^{j}}
&\tableeqn{-i(6\lsp\cOJOA+2\lsp\cOJOC-3\lsp\cOJOD+2\lsp\cOJOF)}
&\tableeqn{
}
&\tableeqn{
}
\\\hline
&$\cQObTQbO4$
&\tableeqn{\invI12\lsp\invI32\lsp(\invI23)^2\lsp\invJ231(\invI13)^{j-2}}
&\tableeqn{-\frac{3i}2\lsp\cOJOF}
&\hfill$\diagup$\hfill\lsp
&\hfill$\diagup$\hfill\lsp
\\\hline
&$\cQObTQbO5$
&\tableeqn{\invI12\lsp\invI31\lsp\invI23\lsp\invJ132\lsp(\invI13)^{j-1}}
&\tableeqn{\begin{aligned}
			&4i\qb\lsp\cOJOB-2i(\qb-1)\lsp\cOJOC\\
			&\;-i(2\qb-3)\lsp\cOJOF
			\end{aligned}}
&\tableeqn{
}
&\hfill$\diagup$\hfill\lsp
\\\hline
&$\cQObTQbO6$
&\tableeqn{\invI12\lsp\invI21\lsp\invJ132\lsp\invJ123\lsp(\invI13)^{j-1}}
&\tableeqn{\frac{3i}2(\cOJOC+\cOJOF)}
&\tableeqn{
}
&\hfill$\diagup$\hfill\lsp
\\\hline
&$\cQObTQbO7$
&\tableeqn{\invI12\lsp\invI21\lsp\invI23\lsp\invI32\lsp(\invI13)^{j-1}}
&\tableeqn{i(\cOJOC+2\lsp\cOJOF)}
&\tableeqn{
}
&\hfill$\diagup$\hfill\lsp
\\\hline
&$\cQObTQbO8$
&\tableeqn{(\invI12)^2\lsp(\invI23)^2\lsp\invJ123\lsp\invJ231(\invI13)^{j-3}}
&\tableeqn{0}
&\hfill$\diagup$\hfill\lsp
&\hfill$\diagup$\hfill\lsp
\\\hline
&$\cQObTQbO9$
&\tableeqn{(\invI12)^2\lsp(\invI23)^2\lsp\invI31\lsp(\invI13)^{j-2}}
&\tableeqn{i(\qb-2)\lsp\cOJOF}
&\hfill$\diagup$\hfill\lsp
&\hfill$\diagup$\hfill\lsp
\\\hline
&$\cQObTQbO{10}$
&\tableeqn{(\invI12)^2\lsp\invI21\lsp\invI23\lsp\invJ123\lsp(\invI13)^{j-2}}
&\tableeqn{-\frac{3i}2\lsp\cOJOF}
&\hfill$\diagup$\hfill\lsp
&\hfill$\diagup$\hfill\lsp
\\\hline
\end{tabularx}
\caption{Expansion of the supersymmetric correlator in the component
$\langle (Q\Ob) T (\Qb O)\rangle$ when $Q\Ob$ has spin $(\frac12,\frac12 j)$. The unbarred entries in the last two columns can be obtained by setting the absent coefficients to zero (see caption of \tablename~\ref{tab:shortening}).}
\label{tab:matchingQObTQbO}
\end{table}}{nonsusy}

\section{The averaged null energy condition}
\label{sec:ANEC}

\par
Following \cite{Cordova:2017dhq, Zhiboedov:2013opa} we define the state
$|\psi\rangle$ of \eqref{eq:ANEC} by acting with some operator $O(x,\eta,\etab)$ on the CFT vacuum $|0\rangle$ and taking the Fourier transform in order to give the state a definite momentum,\footnote{Due to translation invariance, Fourier transforming in both states will lead to an overall $\delta^4(q_1+q_3)$. We simply set $q_3 = - q_1 = q$ and drop the delta function.} which for our purposes we can set to $q^\mu=(1 , \boldsymbol{0})$. Then we multiply by $(x^+)^2/16$ and send $x^+\to \infty$ to simplify the computations. Lastly we need to specify a polarization, but using the auxiliary spinors $\eta$ and $\etab$ we can obtain all possible polarizations at once.
\par
The ANEC integral breaks rotation invariance to an $\mathrm{SO}(2)$
generated by ${\sigma^{12}}_\alpha^{\phantom{\alpha}\beta}$ and
${\sigmab^{12}}^\alphad_{\phantom{\alphad}\betad}$ in the respective
representations. Under a $\varphi$ rotation of this subgroup, fundamental
spinors with a lower index transform as follows:
\eqn{
\left(\begin{array}{l}a\\b\end{array}\right)_\alpha\; \longrightarrow\;
\left(\begin{array}{l}a\,e^{-i\varphi/2}\\b\,e^{i\varphi/2}\end{array}\right)_\alpha \,,\qquad
\left(\begin{array}{l}\bar{a}\\\bar{b}\end{array}\right)_\alphad\; \longrightarrow\;
\left(\begin{array}{l}\bar{a}\,e^{i\varphi/2}\\\bar{b}\,e^{-i\varphi/2}\end{array}\right)_\alphad\,.
}[]
This will help us in the following way: in principle, if there are $s$ choices for the polarization of $O$ and $\Ob$ one would have to apply the ANEC integral to each pair of choices, diagonalize an $s \times s$ matrix and require the positivity of each eigenvalue (or equivalently require semidefinite positiveness of an $s\times s$ matrix). This rotational symmetry reduces the matrix to a block diagonal form, making much simpler the study of its positiveness.
\subsection{Operators of spin \texorpdfstring{$(\frac12 j,0)$}{(j/2,0)}}
Let us focus first on the case where $O(x,\eta,\etab)$ has spin $(\frac12 j,0)$. We
can expand the $\eta$'s in the eigenbasis of the $\mathrm{SO}(2)$ spin,
\eqn{
\eta_3^\alpha = \left(\begin{array}{c}m\\p\end{array}\right) \equiv m\, \boldsymbol{\xi}_-^\alpha + p\,\boldsymbol{\xi}_+^\alpha\,,\qquad \etab_1^\alphad =\left(\begin{array}{c}\bar{p}\\\bar{m}\end{array}\right) \equiv \bar{p}\,\bar{\boldsymbol{\xi}}_+^\alphad + \bar{m}\, \bar{\boldsymbol{\xi}}_-^\alphad\,,
}[eq:polarizations13]
where the redundancy $\boldsymbol{\xi}_\pm = \bar{\boldsymbol{\xi}}_\mp$ has been introduced for convenience. The stress tensor is instead polarized along the null geodesic $u^\mu$, which is translated to
\eqn{
\eta_2^\alpha = \boldsymbol{\xi}_-^\alpha\,,\qquad \etab_2^\alphad = \bar{\boldsymbol{\xi}}_+^\alphad\,.
}[eq:polarizations2]
Now we can perform the ANEC integral~\eqref{eq:ANEC} with the prescriptions defined above on an arbitrary three-point function $\mathsf{t}_{\Ob T O}$.\footnote{The conventions are\[
x^+ = x^0 + x^3 = \boldsymbol{\xi}_- \lsp \rmx\, \bar{\boldsymbol{\xi}}_+\,,\qquad
x^- = x^0 - x^3 = \boldsymbol{\xi}_+ \lsp \rmx\, \bar{\boldsymbol{\xi}}_-\,,\qquad x^2 = - x^+x^- + \vec{x}^2_\perp\,.
\]} We define $x_{13} = x$, $x_{23} = y$ and
\eqn{
\cA[\mathsf{t}_{\Ob T O}] \equiv \int_{-\infty}^\infty \di y^-\,\lim_{y^+\to\infty} \frac{(y^+)^2}{16}\int_{\R^4}\di^4x\,e^{-ix^0}\,\mathsf{t}_{\Ob T O} (x,y;\etab_1,\eta_2,\etab_2,\eta_3)\Big|_{\substack{
\etab_1,\eta_3\to\eqref{eq:polarizations13}\hfill\\
\etab_2,\eta_2 \to \eqref{eq:polarizations2}\hfill}}
\,.
}[]

In order to enforce the correct ordering, the integral in $y^-$ must be
supplemented with the appropriate $i\epsilon$ prescription, namely $y^0 \to y^0 - i \epsilon$ and $x^0 \to x^0 - 2 i \epsilon$. The integrals and the limit $y^+\to \infty$ remove all dependence on the points $x,y$. The result is therefore a polynomial in the variables $p,m,\bar{p}$ and $\bar{m}$. The same considerations apply for the norm of the state, which is computed by Fourier transforming the two-point function
\eqn{
\cF[\mathsf{n}_{\Ob O}] \equiv \int_{\R^4} \di^4x\,e^{-ix^0}\,\mathsf{n}_{\Ob O}(x;\etab_1,\eta_3)\Big|_{
\etab_1,\eta_3\to\eqref{eq:polarizations13}}\,.
}[]

The restrictions imposed by $\mathrm{SO}(2)$ invariance imply that only
certain terms can appear, i.e.\
\eqn{
\cA[\mathsf{t}_{\Ob T O}] = \sum_{s = 0}^j \mathcal{A}_s[\mathsf{t}_{\Ob T O}] \,(p\bar{m})^s(m\bar{p})^{j-s}\,,\qquad
\cF[\mathsf{n}_{\Ob O}] = \sum_{s = 0}^j \mathcal{F}_s[\mathsf{n}_{\Ob O}] \,(p\bar{m})^s(m\bar{p})^{j-s}\,.
}[eq:defAandF]
Each coefficient of this polynomial corresponds to a different choice for the polarizations of $O$ and $\Ob$, therefore the polarization matrix is diagonal and the ANEC states
\eqn{
\anecE{\Delta}{j,0}{s}\equiv \frac{\cA_s[\mathsf{t}_{\Ob T O}]}{\cF_s[\mathsf{n}_{\Ob O}]} \geq 0\,,\qquad \mathrm{for} \;\,s = 0,\ldots, j\,.
}[]
The integrals have been computed explicitly for some values of $j$ in
\cite{Cordova:2017dhq}. Here we provide a general formula, whose proof can
be found in \appendixname~\ref{app:proofEj0}:
\eqna{
\anecE{\Delta}{j,0}{s}&= \frac{3\pi \lsp (-i)^{j}}{8} \frac{(\delta -1)(\delta+j)}{(\delta+j-s-1)_3}\left(\cObTO1+
\frac{j-s}{j}\frac{\delta+j-1}{\delta+j-s-2}\lsp\cObTO2 \right.+\\&
\hspace{5cm}+\left. \frac{(j-s-1)_2}{(j-1)_2}\frac{(\delta-j-2)_2}{(\delta+j-s-3)_{2}}\lsp\cObTO3
\right)\,,
}[eq:genformula]
where $\delta = \Delta - \frac12j -1$ and $(a)_n = \Gamma(a+n)/\Gamma(a)$ is the
Pochhammer symbol. See \tablename~\ref{tab:WIObTO} for the meaning of the
three-point function coefficients. For the special cases $j=0,1$ it suffices to set to zero the absent coefficient(s). Note that \eqref{eq:genformula} is real because the coefficients $\cObTO{i}$ are purely real (resp. imaginary) if $j$ is even (resp. odd).

\subsection{ANEC on a superposition of states}
In the previous subsection the operator $O$ could have been either the
superconformal primary or the first superdescendant $QO^\pm$. However,
these operators mix with each other, i.e.\ the three-point function $\langle (\Qb\Ob^+)T(QO^-)\rangle$ is nonzero. This means that we can impose an even stronger constraint by demanding positivity on the general superposition
\eqn{
| \psi\rangle = \frac{v \lsp(QO^+)|0\rangle}{|\langle(\Qb\Ob^+)(QO^+)\rangle|^{1/2}} +  \frac{w\lsp(QO^-)|0\rangle}{|\langle(\Qb\Ob^-)(QO^-)\rangle|^{1/2}} \,.
}[]
A similar approach was used in \cite{Cordova:2017zej}. Since $v$ and $w$
can be chosen arbitrarily, the ANEC now becomes a semidefinite-positiveness
constraint on a $2(j+1) \times 2(j+1)$ matrix. Such a matrix can be
decomposed in $j$ blocks of size $2\times 2$ and two $1\times1$ blocks, resulting in
\eqna{
\left(
\begin{array}{ll}
\anecE{\Delta+\tfrac12}{j+1,0}{s+1} & \anecEint{\Delta+\tfrac12}{j\pm1,0}{s}\\
\anecEint{\Delta+\tfrac12}{j\pm1,0}{s}&\anecE{\Delta+\tfrac12}{j-1,0}{s}
\end{array}
\right)&\succeq 0 \qquad \mbox{for}\; s = 0,\ldots, j-1\,,\\ \anecE{\Delta+\tfrac12}{j+1,0}{s} &\geq 0\qquad \mbox{for}\;s = 0,j+1\,.
}[]
The diagonal entries have the same expression as \eqref{eq:genformula} with the substitution $\cObTO{i} \to
\cQbObTQOpp{i}$ or $\cObTO{i} \to \cQbObTQOmm{i}$ (see Tables~\ref{tab:matchingQbObTQOpp},~\ref{tab:matchingQbObTQOmm}), together with the appropriate redefinition of $\delta$. The ``interference'' terms $\cE_{\mathrm{int}}$ are defined as follows:\footnote{The definition of $\cA_s$ for the interference correlator is similar to \eqref{eq:defAandF} with the difference that we pick up the term $\bar{m}\bar{p}(p\bar{m})^s(m\bar{p})^{j-s-1}$ for $\langle(\Qb\Ob^+)T(QO^-)\rangle$ and $mp(p\bar{m})^s(m\bar{p})^{j-s-1}$ for its conjugate.}
\eqn{
\anecEint{\Delta+\tfrac12}{j+1,0}{s}=\anecEint{\Delta+\tfrac12}{j-1,0}{s}\equiv
\frac{\cA_s[\mathsf{t}_{(\Qb\Ob^+)T(QO^-)}]}{\big(\cF_{s+1}[\mathsf{n}_{(\Qb\Ob^+)(QO^+)}]\lsp\cF_{s}[\mathsf{n}_{(\Qb\Ob^-)(QO^-)}]\big)^{1/2}}\,.
}[]
Following steps similar to the ones illustrated in \appendixname~\ref{app:proofEj0} one can prove the general formula
\eqna{
\anecEint{\Delta+\tfrac12}{j\pm1,0}{s} &= \frac{3\pi (-i)^{j-1}}{16}
\sqrt{\frac{\delta  (s+1) (j-s)}{j(j+1) (\delta +j+1)}} \frac{(\delta+j-1)_3}{(\delta+j-s-2)_4}\times
\\&\hspace{4cm}\times\left(\frac{\delta+j-s-2}{\delta+j-1}\cQbObTQOpm1 + \frac{j-s-1}{j-1}\cQbObTQOpm2\right)\,,
}[]
where the coefficients $\cQbObTQOpm{i}=\cQbObTQOmp{i}$ are defined in
Tables~\ref{tab:matchingQbObTQOpm}, \ref{tab:matchingQbObTQOmp} and
$\delta = \Delta_{QO} - \frac12 j - \frac32$. Here $\Delta_{QO} = \Delta + \frac12$ is the dimension of the superdescendant. The polarization $s$ takes values from $0$ to $j-1$.

\subsection{Operators of spin \texorpdfstring{$(\frac12
j,\frac12)$}{(j/2,1/2)}}

The only difference when considering more general $\mathrm{SO}(1,3)$
representations is that the polarization matrix will not be diagonal. This
means that the ANEC will not be a set of simple inequalities but rather
semidefinite positiveness constraints. In the $(\frac12 j,\frac12)$ case we further have to specify the polarizations $\eta_1$ and $\etab_3$; thus together with \eqref{eq:polarizations13} and \eqref{eq:polarizations2} one has
\eqn{
\eta_1^\alpha = \left(\begin{array}{c} m'\\ p'\end{array}\right) \equiv  m'\, \boldsymbol{\xi}_-^\alpha +  p'\,\boldsymbol{\xi}_+^\alpha\,,\qquad \etab_3^\alphad =\left(\begin{array}{c}\bar{p}'\\\bar{m}'\end{array}\right) \equiv \bar{p}'\,\bar{\boldsymbol{\xi}}_+^\alphad + \bar{m}'\, \bar{\boldsymbol{\xi}}_-^\alphad\,.
}[eq:otherpolarizations13]

The ANEC integral for an arbitrary operator $O$ of spin $(\frac12 j,\frac12)$ takes the form
\eqna{
\tilde\cA[\mathsf{t}_{\Ob T O}] \equiv \int_{-\infty}^\infty \di y^-\,\lim_{y^+\to\infty} \frac{(y^+)^2}{16}\int_{\R^4}\di^4x\,e^{-ix^0}\,\mathsf{t}_{\Ob T O} (x,y;\eta_{1,2,3},\etab_{1,2,3})\Big|_{\substack{
\etab_1,\eta_3\to\eqref{eq:polarizations13}\hfill\\
\etab_3,\eta_1\to\eqref{eq:otherpolarizations13}\hfill\\
\etab_2,\eta_2 \to \eqref{eq:polarizations2}\hfill}}
\,.
}[]
We also define $\tilde\cF[\mathsf{n}_{\Ob O}]$ in a similar way.
The constraints of $\mathrm{SO}(2)$ invariance allow us to express
\eqn{
\tilde\cA[\mathsf{t}_{\Ob T O}] =\sum_{s=0}^{j+1}\sum_{a,b=0}^1\big(\tilde{\cA}_{s}[\mathsf{t}_{\Ob T O}]\big)_{ab}\,
(p\bar{m})^s(m\bar{p})^{j-s} p'\bar{m}'\left(\frac{\bar{p} m'}{\bar{m} p'}\right)^a\left(\frac{m\bar{p}'}{p\bar{m}'}\right)^b
\,,
}[]
and similarly for $\tilde\cF[\mathsf{n}_{\Ob O}]$. The terms for $s=0$ and $s=j+1$ are restricted to, respectively, $a=b=0$ and $a=b=1$. Thus we can see that the polarization matrix is block diagonal with $j$ blocks of size $2\times2$ and two blocks of size $1\times1$. Defining
\eqn{
\big(\anecE{\Delta}{j,1}{s}\big)_{ab} \equiv \frac{(\tilde{\cA}_{s}[\mathsf{t}_{\Ob T O}])_{ab}}{\big((\tilde{\cF}_{s}[\mathsf{n}_{\Ob O}])_{aa}\lsp\tilde{\cF}_{s}[\mathsf{n}_{\Ob O}])_{bb}\big)^{1/2}}\,,
}[]
the positivity constraints are
\eqna{
& \anecE{\Delta}{j,1}{s} \succeq 0\,,\qquad \mathrm{for}\;\, s=1,\ldots, j\,,\\
& \anecE{\Delta}{j,1}{s} \geq 0\,,\qquad \mathrm{for}\;s= 0, j+1\,.\\
}[]
In the next subsection we will explain how to implement a numerical study
of this system of inequalities. We obtained a general formula for
$\anecE{\Delta}{j,1}{s}$ as well---unfortunately, however, the expression is too unwieldy to be reported here. In \appendixname~\ref{app:proofEj0} we briefly explain how to obtain it.

\subsec{The ANEC as a semidefinite programming problem}[ANECsemidef]

Imposing semidefinite positiveness on a symmetric matrix is a well known
problem for which there exist algorithms that go under the name of \emph{semidefinite programming}. We will make use of the implementation realized by the software \href{https://github.com/davidsd/sdpb}{\texttt{sdpb}}~\cite{Simmons-Duffin:2015qma}, which was developed for the numerical bootstrap approach for the study of CFTs~\cite{Poland:2018epd}, but is general purpose enough to work for our problem too.\par
In general we need to solve a system of inequalities
\eqn{
\anecE{\Delta}{j,\jb}{s} \succeq 0\,,\qquad \mathrm{for}\;\,s = 0, \ldots,  j+\jb\,,
}[]
where $\anecE{\Delta}{j,\jb}{s}$ is a symmetric $m_s\times m_s$ matrix with $m_s =
\min\{j,\,\jb,\,s,\,j+\jb-s\}+1$. The matrices $\cE$ will depend on $N$
arbitrary three-point function coefficients (given by
\tablename~\ref{tab:final}) plus an inhomogeneous part which is fixed by
the Ward identities. Dropping the $\Delta$ and $(j,\jb)$ labels for brevity
one has
\eqn{
\cE[s] = \cE^{(0)}[s] + \sum_{n=1}^{N} \lambda_n\,\cE^{(n)}[s] \succeq 0\,,\qquad \mathrm{for}\;\,s = 0\llsp, \ldots, \lsp j+\jb\,.
}[eq:dualsdp]
This is known as the \emph{dual} formulation of a semidefinite problem. We are interested in studying the feasibility of \eqref{eq:dualsdp}. The algorithm we used only terminates when either a solution
$\lambda_n$ is found, or when a numerical threshold for the internal
computations\footnote{Called \texttt{--maxComplementarity}.} is exceeded.
For our purposes, a problem that terminates for the latter condition is
considered to have no solution.  This means that our ANEC-disallowed points
are not disallowed in a mathematically rigorous way. We expect this to not
have any practical consequences.\footnote{In principle there is also a way to mathematically prove that no solutions exist by providing a certificate of infeasibility~\cite{2011arXiv1108.5930K}. By using \cite{Laurent2009,Kojima2003,Scherer2006} this amounts to finding a solution of another (larger) semidefinite problem.}

\subsection{Details on ANEC bounds: nonsupersymmetric case}
\label{sec:nonsusy}

Let us briefly review the results obtained in \cite{Cordova:2017dhq} and
prove a few results for generic values of $j$. First let us consider
conformal primaries in the $(\frac12 j, 0)$ Lorentz representation.
The ANEC condition is expressed by the formula \eref{eq:genformula}, where
the coefficients $ D_i$ are given in Table~\ref{tab:WIObTO}. In particular, one can take $\hat D_1=-i^j D_1$ to be the only independent real coefficient. By choosing the value $s=0$ and $s=j$ in \eref{eq:genformula} and restricting to the case $j>2$ for simplicity
we obtain
\eqn{
(\delta -1) \big((\pi ^2 \hat D_1-4) \delta +j (\pi ^2 \hat D_1+2 \delta
-6)+2 j^2+4\big) \geq 0\,, \qquad \hat D_1\geq 0\,,
}[eq:nonsusyJ0]
where $\delta\geq0$ represents the distance from the unitarity bound. It is
straightforward to verify that the above conditions cannot be
simultaneously satisfied unless $\delta\geq1$.

\begin{figure}[t]
\centering
\includegraphics{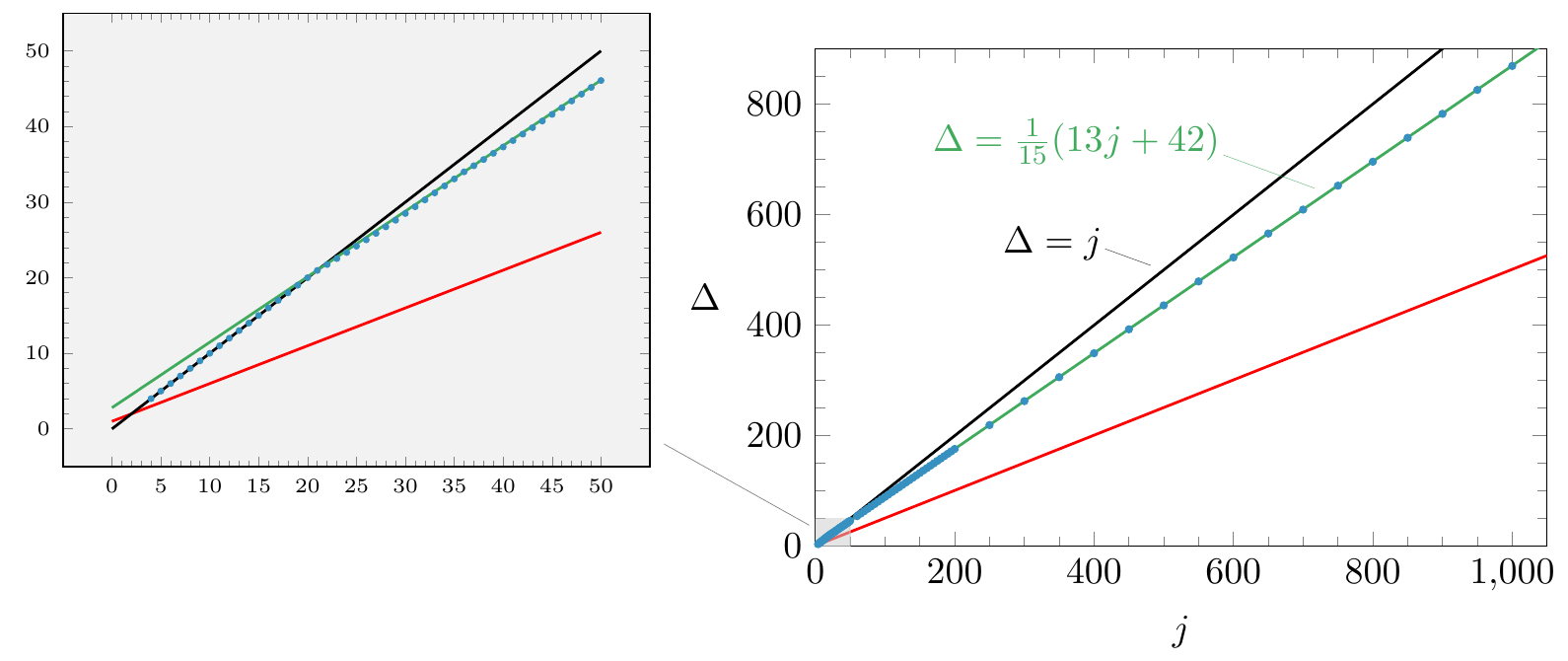}
\caption{Lower bounds on the conformal dimension $\Delta$ as a result of
  the ANEC for primaries transforming in the $(\frac12 j,0)$
  Lorentz representation. Each point is the result of a bisection in
  $\Delta$. The red line is the unitarity bound, $\Delta = \frac12 j + 1$.
  The black line corresponds to the conjecture of~\cite{Cordova:2017dhq},
  $\Delta=j$, and the green line gives an approximate behavior of the
  bound valid above $j=20$.
}\label{fig:nonsusyJ0}
\end{figure}

By considering all polarizations we can obtain stronger bounds at the price
of fixing the value of $j$, for instance by using the function
\texttt{Reduce} of \emph{Mathematica}. We show our results in
\figurename~\ref{fig:nonsusyJ0} up to $j=10^{3}$. Although the bound
initially agrees with the conjecture of \cite{Cordova:2017dhq}, it departs
from it for $j\geq 21$ and follows a different pattern which is well fitted
by the expression $\Delta=\frac12j+1+\delta \geq  \frac{1}{15}(13\lsp
j+42)$. It would be tempting to assign a meaning to the kink at $j\sim 21$,
but the explanation might simply reside in the fact that, going to large
values, the integer nature of $j$ becomes less and less important and new
solutions for $\hat D_1$ become available.

\begin{figure}[t]
\centering
\includegraphics{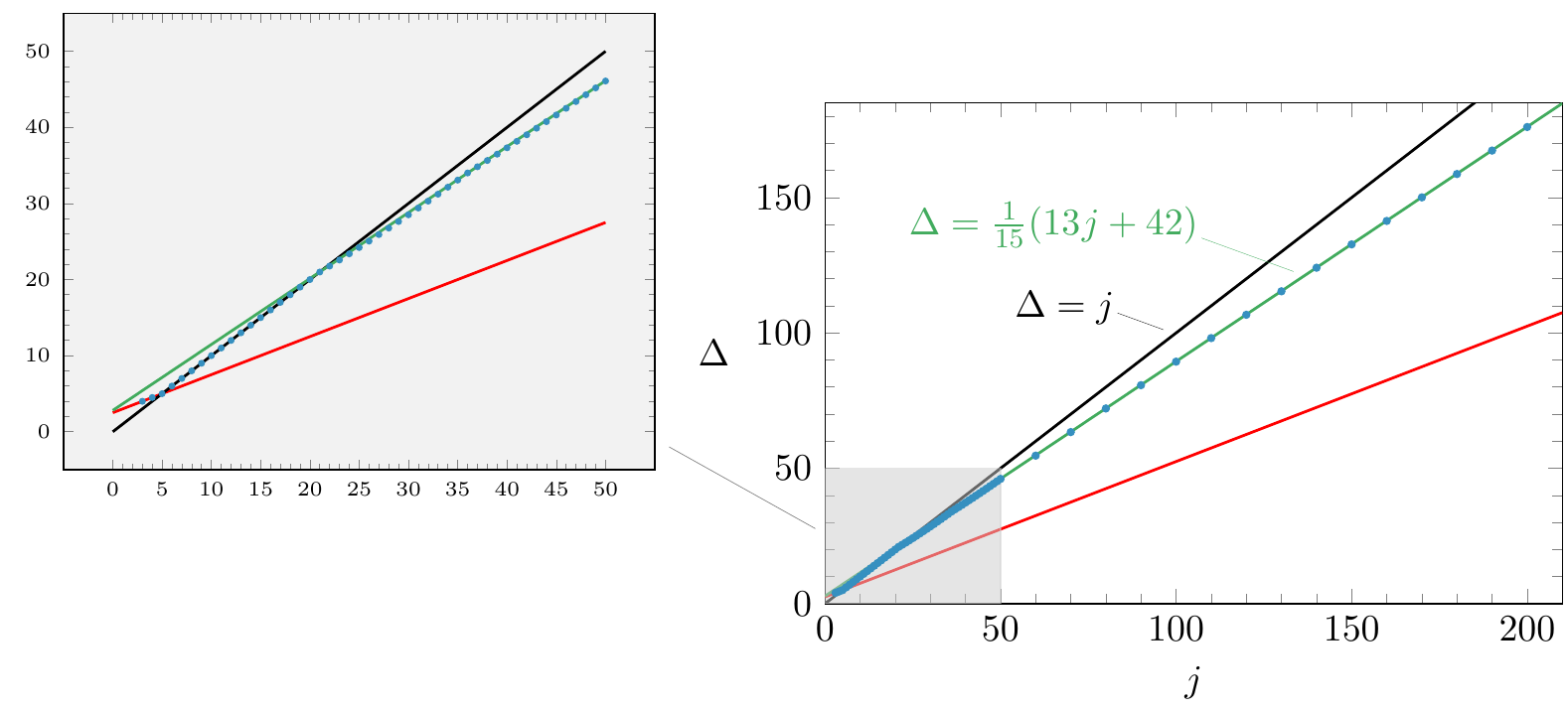}
\caption{Lower bounds on the conformal dimension $\Delta$ as a result of
  the ANEC for primaries transforming in the $(\frac12 j,\frac12)$
  Lorentz representation. Each point is the result of a bisection in
  $\Delta$. The red line is the unitarity bound, $\Delta = \frac12 j +
  \frac52$. The operators for $j\leq 5$  lie on the red line. The black
  line corresponds to the conjecture of~\cite{Cordova:2017dhq}, $\Delta=j$, and the green line gives an approximate behavior of the
  bound valid above $j=20$.
}\label{fig:nonsusyJ1}
\end{figure}

Let us now move to the case of conformal primaries in the $(\frac12 j,
\frac12)$ representation. The procedure to obtain the general formula is
described in Appendix~\ref{sec:formulaJ1}. After imposing the Ward
identities, whose solution is reported in Table~\ref{tab:WIQObTQbO}, one is
left with four independent three-point function coefficients $H_i$. In
order to systematically address the feasibility of the ANEC we translated
the linear matrix inequality into a semidefinite problem as discussed in
the previous subsection. We found agreement with the results of
\cite{Cordova:2017dhq} for $j\leq 7$ and extended the bounds up to $j=50$.
A lower bound on $\Delta$ as a function of $j$ is shown in
\figurename~\ref{fig:nonsusyJ1}: again we observe that for $j>21$ the
bounds departs from the conjecture $\Delta\geq j$ of
\cite{Cordova:2017dhq} and closely follows the bound
$\Delta\geq \tfrac1{15}(13j+42)$ instead.

In the case of conserved operators the problem simplifies considerably: only two coefficients remain independent\footnote{The relation imposed by conservation of the operator $O$ can be easily computed using the package \href{https://gitlab.com/bootstrapcollaboration/CFTs4D}{\texttt{CFTs4D}}.} and we can easily prove that conserved currents cannot exist for $j>5$.
For instance, we can take $\hat H_{9,10}= -i^{j+1} H_{9,10}$ to be the two
independent real coefficients. By considering the eigenvalues of matrices with  $s=j-3,\ldots,j$ and the condition at $s=j+1$, we obtain the following set of inequalities:
\begin{eqnarray}
\hat H_{10} \geq 0\,,\qquad  3  \hat H_ 9+\frac{18}{\pi^2 }\frac{j-1}{j+1}
\leq \hat H_{10} \frac{2 j+1}{j-1}\,, \nonumber \\
\hat H_9 \leq \frac23 \hat H_{10} \,,\qquad 3  \hat H_9 +\frac{12}{\pi^2}  \geq   2 \hat H_{10} \frac{j+1}{j-1}\,.
\end{eqnarray}
One can immediately check that the above conditions admit a solution only
for $j\leq 5$, corresponding to the cases when conserved currents can be
constructed in free theories. Interestingly, for the boundary case $j=5$ the solution to the ANEC is unique:
\eqn{
  \hat H_9 = -\frac{4}{\pi^2}\,,\qquad\hat H_{10}=0\,.
}[eq:nonsusy51]

\subsection{Details on ANEC bounds: supersymmetric case}
In the supersymmetric case the analysis follows the same steps as before,
except that now one needs to combine multiple conditions. Let us discuss
some of the results presented in the introduction. We first start from
a multiplet whose zero component transforms in the $(\frac12 j,0)$
representation and satisfies the $[L,\overbar{B}]$ shortening condition.
These are the generalizations to $j>1$ of the usual chiral scalar and
gauge-invariant spin-$\frac12$ multiplets. In this case $\bar{q}=0$ and
$q=\Delta$. The multiplet contains only four conformal primaries: $O$,
$QO^{\pm}$ and $Q^2 O$. In this work we only consider the first three. As
discussed in \sectionname~\ref{sec:shortening} the superspace three-point
function does not have any free parameters. Let us consider, then, the ANEC
applied to the superprimary only. The condition is again encoded in
\eref{eq:genformula}, where now the coefficients $D_i$ are related to the
superspace coefficients through the relations in
Table~\ref{tab:matchingObTO}, supplemented by the relations in
Table~\ref{tab:shortening}. The analog of $\hat D_1\geq 0 $ in
\eref{eq:nonsusyJ0} is now simply
\eqn{2 q - 3 j\geq 0\,.}[]
We explicitly checked that including other constraints does not strengthen the bound. This is expected since one can construct chiral operators with $\Delta=\frac32j$ by taking products of free chiral vector multiplets. The bound is therefore optimal.

Let us move to another simple case, namely $[A_1,\overbar{A}_2]$,
corresponding to superprimaries again in the $(\frac12 j,0)$ representation
with $q=\frac12j+1$ and $\bar{q}=1$. This multiplet contains conserved
operators in the $(\frac12(j+1),\frac12)$ and, due to the results of the
previous subsection, we can immediately conclude that $j\geq 4$. It turns
out, however, that $j=4$ is excluded since the values $\hat H_9$ and $\hat
H_{10}$ fixed by supersymmetry do not satisfy \eref{eq:nonsusy51}. Smaller
values of $j$ must be consistent since these operators appear in the
decomposition of extended supersymmetry multiplets in the free limit.

All other bounds found in this work were obtained with a numerical
approach. For completeness we collect here all the conditions we imposed in
the most complicated case $[L,\overbar{L}]$. In simpler cases some of them
do not appear since the corresponding superdescendant is absent. At the
same time, the correct three-point function coefficient relations must be
imposed.  Given an $[L,\overbar{L}]$ supermultiplet with a superprimary
transforming in the $(\frac12 j,0)$ representation and $q\geq \frac12j+1$,
$\bar{q}\geq 1$, the ANEC can be satisfied if there exist real coefficients
$\hat{\mathcal C}_k=i^j \mathcal C_k$, $k=2,6$, such that
\begin{align}
&\langle \Ob TO \rangle: \nonumber\\
&\qquad  \mathcal E[\Delta, (j,0); s] \geq 0  \,,\qquad \mathrm{for}\;\,s = 0, \ldots,  j\,, \nonumber\\
&\langle (\Qb\Ob) T( QO) \rangle: \nonumber\\
\vspace{1em}
&\qquad\!\!\!\begin{array}{ll}
\left(
\begin{array}{lr}
\anecE{\Delta+\frac12}{j+1,0}{s+1} &\qquad \anecEint{\Delta+\frac12}{j\pm1,0}{s}\\
\anecEint{\Delta+\frac12}{j\pm1,0}{s}&\anecE{\Delta+\frac12}{j-1,0}{s}
\end{array}
\right)\succeq 0 \qquad & \mbox{for}\; s = 0,\ldots, j-1\,,\\ \anecE{\Delta+\frac12}{j+1,0}{s} \geq 0 \qquad & \mbox{for}\;s = 0,j+1\,,
\end{array}
\nonumber\\
&\langle ( Q\Ob) T( \Qb O) \rangle: \nonumber\\
\vspace{1em}
 & \qquad\anecE{\Delta+\tfrac12}{j,1}{s} \succeq  0  \,,\qquad \mathrm{for}\;\,s = 0,\ldots,j+1\,.
\end{align}
As usual we defined $\Delta=q+\qb$. Whenever the above system of conditions does not admit a solution, we
conclude that the corresponding supersymmetry multiplet cannot exist in a
local unitary SCFT.

\section{Bounds on extended supersymmetry multiplets}
\label{sec:extendedsusy} \label{sec:extendedSUSY} \subsection{Conventions}
The aim of this section is to constrain the superconformal multiplets of
theories with $\cN > 1$ supersymmetry by decomposing them into $\cN=1$
multiplets. This approach does not make use of the additional linear
relations among the three-point function coefficients and thus may not yield optimal bounds.
Following \cite{Cordova:2016emh}, we will denote $\mathcal N=2$
supermultiplets as $\cX_L\overbar\cX_R[j,\jb]_\Delta^{(\mathcal R, r)}$,
where $(\mathcal R,r)$ are the quantum numbers under the $
\mathfrak{su}(2)\lsp\oplus\lsp\llsp \mathfrak{u}(1) $ algebra, while we will denote $\mathcal
N=4$ supermultiplets as
$\cX_L\overbar\cX_R[j,\jb]_\Delta^{(p_1,p_2,p_3)}$, where
$p_1,p_2$ and $p_3$ are the Dynkin labels of the $\mathfrak{su}(4)$ algebra representation $[p_1,p_2,p_3]$, for which we use the conventions of
\cite{Dolan:2002zh}. As in previous sections, the left/right shortening can
take values $\cX_{L,R}=L,A_1,A_2,B_1$.

We define the supercharges to transform under the $\mathfrak{u}(1)$ R-symmetry of the $\cN \neq 4$ superalgebra as
\eqn{
[r_{\cN}, Q^I_\alpha ] = - Q^I_\alpha\,,\qquad
[r_{\cN}, \Qb_{I\alphad} ] = \Qb_{I\alphad}\,.
}[]
We consider for any $\cN$ the superalgebra generated by $Q^1_\alpha$ and $\Qb_{1\lsp\alphad}$. The embedding of the $\cN=1$ $\mathfrak{u}(1)$ R-charge in the larger R-symmetry group is
\eqn{
\begin{aligned}
\cN=2\,:\qquad && R \equiv r_{\cN=1} &= - \tfrac43 R_3 + \tfrac13 r_{\cN=2}\,, \\
\cN=4\,:\qquad && R \equiv r_{\cN=1} &= - \tfrac13(3 H_1+2H_2+H_3)\,, \\
\end{aligned}
}[]
where $H_i$ is the Cartan generator associated to the $i$-th Dynkin label in $[p_1,p_2,p_3]$.
The generator $R_3$ is the $\mathfrak{su}(2)$ Cartan in units of $\tfrac12$ ($R_3 = -\tfrac12\cR,\ldots,\tfrac12\cR$). Consistently with the rest of the paper, $R$ is the $\cN=1$ R-charge. We will also abbreviate $r \equiv r_{\cN=2}$.

\subsection{\texorpdfstring{$\cN=2$}{N=2}}
\label{sec:N=2}

Let us start by considering the so-called ``exotic chiral primaries,''
namely the $L\overbar{B}_1[j;0]^{(0,r)}_{\Delta}$ multiplets, with $\Delta
= \tfrac12r$.\footnote{Denoted $\overbar{\cE}_{\frac{r}2(j,0)}$ in
\cite{Dolan:2002zh}.} The bound on chiral multiplets \eqref{eq:chiralbound}
for the $\cN=1$ subalgebra generated by $Q_\alpha^{1}$, applied to the
chiral superprimary
$Q^{2}_{(\alpha_1}\CO^{\text{(exotic)}}_{\alpha_2\ldots\alpha_{j+1})}$
implies that
\eqn{\Delta + \tfrac12\geq\tfrac32(j+1)
\quad\Rightarrow\quad\Delta \geq \tfrac{3}{2}j + 1\,.}[exoticBound]
The unitarity bound is $\Delta \geq \tfrac{1}{2}j + 1$, and so we see that
the ANEC bound is stronger for $j>0$.  \par A similar argument can be made
on operators with nonzero $\mathfrak{su}(2)$ R-charge
$L\overbar{B}_1[j;0]_{\Delta}^{(\cR,r)}$, where $\Delta = \cR + \tfrac12r$
and $\cR$ is in integer units. We considered several values of $\cR$ and
performed the decomposition into $\cN=1$ multiplets. Imposing
\eqref{eq:chiralbound} on each of the chiral multiplets that appear yields
the following pattern (which we conjecture to be true for arbitrary $\cR$):
\eqn{ r \geq 3 j + 2 - 2\cR\qquad \Rightarrow \qquad \Delta \geq \tfrac32j
+ 1\,.  }[] This is stronger than unitarity ($r \geq j+2$) for $j > \cR$.
As a consequence, short multiplets of the form $A_\ell
\overbar{B}_1[j;0]^{(\cR,r)}_{\Delta}$ are only allowed for $j \leq \cR$.
\par The multiplets $A_1 \overbar{B}_1[j;0]^{(1,j+2)}_\Delta$ and $A_1
\overbar{A}_2[j;0]^{(0,j)}_\Delta$ with $\Delta = \tfrac12j+2$ are absent
from any local SCFT for $j>2$. This is a consequence of the presence of an
$A_1\overbar{A}_2[j+1;0]$ multiplet in their $\cN=1$ decomposition, which
we have shown to be forbidden by the ANEC when $j+1 > 3$.  \par We also
considered long multiplets $L\overbar{L}[j;0]^{(\cR,r)}_\Delta$ for some
values of $\cR$. Calling $\delta$ the difference of their dimension and
their unitarity bound,
\eqn{ \delta = \Delta - 2 - j - \cR + \tfrac12 r\,,
}[]
and calling $f(R,j)$ the separation between the unitarity and the ANEC
bound in \figurename~\ref{fig:LLbRDelta}, we find the following pattern
\eqn{ \delta \geq f\big(\tfrac13(r+1),j+1\big) - \cR\,.  }[]

\subsection{\texorpdfstring{$\cN=4$}{N=4}}
We considered a few short multiplets and found no constraints from the
ANEC. Interestingly, $B_1\overbar{B}_1[0;0]^{(1,0,1)}_2$ contains a chiral multiplet that saturates \eqref{eq:chiralbound}, namely
\eqn{
B_1\overbar{B}_1[0;0]^{(1,0,1)}_2 \supset L\overbar{B}_1[2;0]^{(2)}_3\,.
}[]
\par
The simplest long multiplet is the Konishi multiplet $L\overbar{L}[0;0]^{(0,0,0)}$. In its $\cN=1$ decomposition we find a long multiplet of spin $(\tfrac32,0)$ and R-charge $1$ with dimension $\Delta_\mathrm{Konishi} + \tfrac32$. In terms of the $Q^{1}_\alpha$ subalgebra, calling $\phi$ the Konishi operator, one has
\eqn{
\cO_{\alpha_1\alpha_2\alpha_3} = \varepsilon_{1IJK} \lsp Q^I_{(\alpha_1}Q^J_{\alpha_2}Q^K_{\alpha_3)}\lsp \phi\,.
}[eq:QQQKon]
Since in perturbation theory one can compute $\Delta_\mathrm{Konishi} = 2 + O(g^2)$, we see that the ANEC and the unitarity bound for $\cN=1$ long multiplets of spin $(\tfrac32,0)$ are saturated.
\par
More generally, we checked some cases of long multiplets
$L\overbar{L}[j;0]^{(p_1,p_2,p_3)}$, namely those with Dynkin labels $[p_1,p_2,p_3] = [0,0,0]$, $[0,2,0]$ and $[1,0,1]$. Calling $\delta$ the
difference of their dimension and their unitarity bound,
\eqn{
\delta = \Delta - 2 - j - \tfrac12(3p_1+2p_2+p_3)\,,
}[]
and calling $f(R,j)$ the separation between the unitarity and the ANEC bound in \figurename~\ref{fig:LLbRDelta}, we find
\eqna{
[0,0,0]\,&:\qquad \delta	\geq f\big(\tfrac43,j+2\big) - 2\,,\\
[0,2,0]\,&:\qquad \delta	\geq f\big(\tfrac73,j+3\big) - 4\,,\\
[1,0,1]\,&:\qquad \delta	\geq f\big(\tfrac73,j+3\big) - 4\,.\\
}[]

\section{Conclusions and outlook}\label{sec:outlook}
In this paper we studied effects of the ANEC on the operator spectrum of
CFTs. In particular, we showed that the ANEC imposes lower bounds on
operator dimensions that are stronger than unitarity bounds.  Our
considerations were mostly limited to the case of $\mathcal N=1$
superconformal multiplets whose superconformal primaries transform in the
$(\tfrac12j,0)$ representation of the Lorentz group. This suffices to show
that the unitarity bounds are typically suboptimal to the ANEC bounds.

Our methods apply in more general situations, with or without
supersymmetry. It would be of great value to obtain an educated guess for
the ANEC bound on multiplets whose superconformal primaries transform in
the general $(\tfrac12j,\tfrac12\jb)$ representation. In this respect, the
techniques presented here to compute the ANEC integral in closed form and
the usage of semidefinite programming will considerably simplify  the
analysis.

These ideas can also be generalized to extended supersymmetry, in
particular $\cN=2$. In principle it is possible to carry out a similar
analysis for the three-point functions in $\cN=2$ superspace with a
formalism similar to the one used in this paper and using results
of~\cite{Kuzenko:1999pi,Ramirez:2016lyk}. One of the motivations behind
pursuing this direction would be to potentially further constrain the
exotic chiral primaries $L\overbar{B}_1[j;0]^{(0,r)}_{r/2}$. These
operators for $j \geq 1$ have been proved to be absent in a very large
class of theories~\cite{Buican:2014qla}. Using the results in
\sectionname~\ref{sec:extendedsusy} we are able to constrain their
dimension to
\eqn{
\Delta_{\mathrm{exotic}} \geq \tfrac{3}{2}j +
1\,.
}[]
It would be interesting to see if  ANEC forbids them in general once the
$\mathcal N=2$ superconformal symmetry is fully taken into account.

In $\cN=2$ one could also investigate the higher-spin version of the ANEC
mentioned in the introduction~\cite{Komargodski:2016gci,Meltzer:2018tnm}.
In a generic CFT it is hard to address such a problem because, unlike the
spin-two case, the dimension of the lowest-twist operator is not fixed and
there are no Ward identities to constrain the three-point function
coefficients. In $\cN=2$ SCFTs, however, there are higher-spin operators
with protected dimensions that are not at the unitarity bound (hence do not
decouple from the theory~\cite{Maldacena:2011jn,Alba:2015upa}). An example
are the $A_1\overbar{A}_1[\ell;\ell]_\Delta^{(R,0)}$ multiplets, with
$\Delta = \ell + 2 + R$ and $R>0$. Clearly the bounds obtained this way
will not be general but will assume that $R$ is the smallest R-charge among
these protected operators and, at spin $\ell$, the unprotected spectrum has
a gap larger than $\ell+2+R$. We leave these questions for future
investigations.

\ack{This work was initiated at the Bootstrap 2018 conference held at
Caltech.  We thank the organizers for creating a stimulating atmosphere,
and the participants for interesting conversations. We especially thank
Clay Cordova for inspiring discussions about his work. We also thank Kenan
Diab, Zohar Komargodski and Leonardo Rastelli for useful comments. Finally,
we thank Ning Su for helpful comments about the performance of
\texttt{sdpb}.  AM and AV are supported by the Swiss National Science
Foundation under grant no.\ PP00P2-163670. AV is also supported by the
European Research Council Starting Grant under grant no.\ 758903. Some of
the computations in this paper were run on the EPFL SCITAS cluster.}

\begin{appendices}
\section{Supersymmetric inversion tensors}\label{app:tensorsI}
Here we list the properties needed to derive equation~\eqref{eq:realityEqANEC}. The order in which they appear is roughly the order in which one needs to apply them. First of all, the explicit definition of the tensors is
\threeseqn{
I_{\mu\nu}(x_{1\bar{2}},x_{\bar{1} 2}) &= \bar{I}_{\nu\mu}(x_{\bar{2} 1},x_{2\bar{1}}) = \frac{\tr(\sigma_{\mu}\tilde\rmx_{\bar{1} 2}\sigma_\nu\tilde\rmx_{\bar{2} 1})}{2\sqrt{{x_{\bar{1} 2}}{\!}^2{x_{\bar{2} 1}}{\!}^2}} = \frac{\tr(\sigmab_{\mu}\rmx_{1 \bar{2}}\sigmab_\nu\tilde\rmx_{2 \bar{1}})}{2\sqrt{{x_{\bar{1} 2}}{\!}^2{x_{\bar{2} 1}}{\!}^2}}\,,
}[]{
I^{i\ib}(x_{1\bar{2}}) &= \frac{i^j}{j!}\frac{(\rmx_{1\bar{2}})_{\alpha_1(\alphad_1}\cdots (\rmx_{1\bar{2}})_{\alpha_j|\alphad_j)}}{{x_{\bar{2}1}}^j}\,,
}[]{
\bar{I}_{\ib i}(x_{\bar{2}1}) &= \frac{(-i)^j}{j!}\frac{(\tilde{\rmx}_{\bar{2}1})^{\alphad_1(\alpha_1}\cdots (\tilde{\rmx}_{2\bar{1}})^{\alphad_j|\alpha_j)}}{{x_{\bar{2}1}}^j}\,.
}[]
The needed properties are
\threeseqn{
I_{\mu\nu}(x,\xb)\bar{I}^{\nu\rho}(-x,-\xb) &= \delta_\mu^\rho\,,
}[]{
I^{i\ib}(x)\bar{I}_{\ib i'}(-x) &= \delta^i_{i'}\,, 
}[]{
I_{\lambda\rho}(x_{1\bar{3}},x_{\bar{1}3})\bar{I}^{\rho\nu}(x_{\bar{3}2},x_{3\bar{2}})I_{\nu\mu}(x_{2\bar{1}},x_{\bar{2}1})  &= I_{\lambda\mu}(-\Xb_1,-X_1)\,.
}[]
The covariance property of the $t$ and its $\lambda\bar{\lambda}$ scaling \eqref{eq:homogeneityEta} imply
\eqna{
&
I^{i_1\ib_1}(x_{1\bar{3}})I^{i_4\ib_3}(x_{1\bar{3}})\,I_{\lambda\nu}(x_{1\bar{3}},x_{\bar{1}3})\lsp
t^{\phantom{\ib_1}\nu}_{\ib_1\phantom{\nu\,}\ib_3}(Z_3) = \\ &\hspace{3cm}= X_1{\!}^3{\Xb_1}{\!}^3{x_{\bar{1}3}}{\!}^3\,{x_{\bar{3}1}}{\!}^3I^{i_1\ib_1}(\Xb_1)I^{i_4\ib_3}(\Xb_1)\,I_{\lambda\nu}(\Xb_1,X_1)\lsp t^{\phantom{\ib_1}\nu}_{\ib_1\phantom{\nu\,}\ib_3}(Z_1)\,.
}[]
The last identities that we need are
\eqn{
{X_1}{\!}^2 = \frac{{x_{\bar{2}3}}{\!}^2}{{x_{\bar{2}1}}{\!}^2 {x_{\bar{1}3}}{\!}^2}\,,\qquad
{\overbar{X}_1}{\!}^2 = \frac{{x_{\bar{3}2}}{\!}^2}{{x_{\bar{3}1}}{\!}^2 {x_{\bar{1}2}}{\!}^2}\,.
}[]

\section{Proof of the general formula}\label{app:proofEj0}
\subsection{Formula for the \texorpdfstring{$(\frac12 j,0)$}{(j/2,0)} case}
In this section we provide a proof of the formula \eqref{eq:genformula} which we reproduce here for convenience:
\eqna{
\anecE{\Delta}{j,0}{s}&= \frac{\cA_s[\mathsf{t}_{\Ob T O}]}{\cF_s[\mathsf{n}_{\Ob O}]} = \frac{3\pi \lsp (-i)^{j}}{8} \frac{(\delta -1)(\delta+j)}{(\delta+j-s-1)_3}\left(\cObTO1+
\frac{j-s}{j}\frac{\delta+j-1}{\delta+j-s-2}\lsp\cObTO2 \right.+\\&
\hspace{6.2cm}+\left. \frac{(j-s-1)_2}{(j-1)_2}\frac{(\delta-j-2)_2}{(\delta+j-s-3)_{2}}\lsp\cObTO3
\right)\,.
}[]
The first step is to realize that the dependence on $j$ and $s$ is entirely coming from the tensors $(\invI13)^{\tilde{\jmath}}$ which appear both in $\mathsf{t}_{\Ob T O}$ at the numerator (with $\tilde{\jmath} = j,j-1,j-2$) and in $\mathsf{n}_{\Ob O}$ at the denominator (with $\tilde{\jmath} = j$). Let us then expand this tensor when the polarizations are replaced as in \eqref{eq:polarizations13},
\eqna{
(\invI13)^{\tilde{\jmath}} &= (\eta_3\rmx\etab_1)^{\tilde{\jmath}}
= \big(m\bar{p}\,x^+ + p\bar{m}\,x^- + m\bar{m}\,\rmx_{-\dot{-}} + p\bar{p}\,\rmx_{+\dot{+}}\big)^{\tilde{\jmath}} \\
&=\sum_{s=0}^{\tilde{\jmath}}\sum_{r=0}^{\min(s,\tilde{\jmath}-s)} \binom{\tilde{\jmath}}{2r}\binom{\tilde{\jmath}-2r}{s-r}\binom{2r}{r}\,(x^-)^{s-r}(x^+)^{\tilde{\jmath}-r-s}(x_\perp^2)^r\,(p\bar{m})^s (m\bar{p})^{\tilde{\jmath}-s}\,.
}[]
We obtained this result by simply doing a double binomial expansion and
using $\rmx_{-\dot{-}}\rmx_{+\dot{+}} = x_\perp^2 \equiv (x^1)^2+(x^2)^2$.
All terms where $\rmx_{-\dot{-}}$ and $\rmx_{+\dot{+}}$ appear with
different powers can be thrown away as they are not $\mathrm{SO}(2)$
neutral and there are no other invariants in the tensor structures that can
compensate for them.\footnote{This statement holds in the $y^+\to\infty$
limit.} The first sum is precisely the sum over polarizations, and so we
can remove it and focus on one $s$ at a time. The second sum, instead, can be extended to $\sum_{r=0}^\infty$ since the binomial coefficients are automatically zero when $r$ is out of bounds. This fact will be useful later on.
\par
This expansion completely takes care of the polarizations of
$\mathsf{n}_{\Ob O}$ and of the structure $\cObTO1$ of $\mathsf{t}_{\Ob T
O}$. For the other two structures it is not hard to see that the terms
$(p\bar{m})^s (m\bar{p})^{\tilde{\jmath}-s}$ of the
$(\invI13)^{\tilde{\jmath}}$ tensor of each structure all contribute to the
same term $(p\bar{m})^s (m\bar{p})^{j-s}$.\footnote{To be more precise
there are contributions also to the terms $(p\bar{m})^{s+a}
(m\bar{p})^{j-s-a}$ ($a=1,2$), but it can be verified that in the limit
$y^+\to \infty$ they are subleading.} Concretely we find
\eqna{
\cA_s[\mathsf{t}_{\Ob T O}] = -\frac{3i\pi}{4}&\int_{\R^4}\di^4x\,e^{-ix^0}\, \sum_{r=0}\,
(x^-)^{s-r-5}(x^+)^{j-r-s-2}(x_\perp^2)^r (x^2)^{1-\Delta-j/2}\times
\\&\hspace{0.5cm}\times\left(
I^{(j)}_{r,s}\lsp(x^-)^2(x^+)^2\lsp \cObTO1
-I^{(j-1)}_{r,s}\lsp x^-x^+x^2\lsp \cObTO2
+I^{(j-2)}_{r,s}\lsp(x^2)^2\lsp \cObTO3\right)\,,
}[]
where
\eqn{
I^{(j)}_{r,s} = \binom{j}{2r}\binom{j-2r}{s-r}\binom{2r}{r}\,.
}[]
Similarly, the denominator has the form
\eqn{
\cF_s[\mathsf{n}_{\Ob O}] =i^j \int_{\R^4}\di^4x\,e^{-ix^0}\, \sum_{r'=0}\,I^{(j)}_{r',s} \,(x^-)^{s-r'}(x^+)^{j-r'-s}(x_\perp^2)^{r'} (x^2)^{-\Delta-j/2}\,.
}[]
The Fourier transforms can be straightforwardly computed using the general formulas
\eqna{
\int_{\R^2}\di^2x_\perp\,(x^2)^a(x_\perp^2)^b &= \frac{\pi\lsp \Gamma(1-a-b)\lsp\Gamma(1+b)}{\Gamma(-a)}(-x^- x^+)^{1+a+b}\,,\\
\int_{\R^2} \di x^+\di x^-\,e^{-i(x^++x^-)/2}\,(x^+)^a(x^-)^b &= \frac{(2\pi)^2 (-i)^{a+b}(-2)^{a+b+2}}{\Gamma(-a)\lsp\Gamma(-b)}\,.
}[eq:FTformulas]
What remains now is to compute the sums in $r$ and $r'$. After some
simplifications all sums can be reduced to the following general form for some $m,n$:\footnote{$(m,n)$ can be $(1,1),(2,1),(3,0)$ or $(4,-1)$}
\eqn{
\Sigma_{m,n}=\sum_{r=0}^\infty\frac{(-1)^r}{r!} \frac{\Gamma\big(\Delta + \lifrac{j}{2}-r-m\big)}{\Gamma(1-r+s)\,\Gamma(j-r-s+n)}\,.}[eq:Sigmasum]
We stress again that even though the upper limit is $\infty$, there are actually only a finite number of nonzero terms. After using the property
\eqn{
\Gamma(X-r) = (-1)^r\frac{\Gamma(X)}{(1-X)_r}
}[]
of the $\Gamma$ function, we can rewrite this sum in the form of a
${}_2F_1$ hypergeometric function evaluated at $1$, for which the explicit expression is known:
\eqna{
\Sigma_{m,n} &= \frac{\Gamma\big(\Delta + \lifrac{j}{2}-m\big)}{\Gamma(1+s)\,\Gamma(j-s+n)}\,{}_2F_1\big(-s,1-j-n+s;1-\Delta-\lifrac{j}{2}+m;\,1\lsp\big)\,\\
&=\frac{\Gamma\big(\Delta + \lifrac{j}{2}-m\big)}{\Gamma(1+s)\,\Gamma(j-s+n)}\,
\frac{\Gamma \big(1-\Delta-\lifrac{j}{2}+m\big)\lsp \Gamma \big(\lifrac{j}{2}+m+n-\Delta \big)}{\Gamma \big(s+1-\Delta-\lifrac{j}{2}+m\big) \lsp \Gamma \big(\lifrac{j}{2}+m+n-\Delta-s \big)}\,.
}
The final result will be expressed in terms of ratios $\Sigma_{m,n}/\Sigma_{1,1}$ which are rational functions of $\Delta,j$ and $s$. It is now straightforward to check that it agrees with the general formula \eqref{eq:genformula}.

\subsection{Formula for the \texorpdfstring{$(\frac12 j,\frac12)$}{(j/2,1/2)} case}
\label{sec:formulaJ1}

In order to obtain a formula for this case we mostly need to follow the
same steps as in the previous subsection, with some minor modifications. The main difference is that the invariants $\invI31$, $\invJ123$ and $\invJ231$ can yield contributions with $\mathrm{SO}(2)$ charge $\pm1$ in the limit $y^+\to\infty$. By looking at \tablename~\ref{tab:matchingQObTQbO} we see that all tensor structures have at most one of these invariant except for $\cQObTQbO8$ which contains two. Since that particular structure is zero in our superspace correlator we will not compute a formula for it. As a consequence we need to expand $(\invI13)^{\tilde{\jmath}}$ keeping also terms of charge $\pm1$. This is easily done as follows:
\eqna{
(\invI13)^{\tilde{\jmath}} &= \big(m\bar{p}\,x^+ + p\bar{m}\,x^- + m\bar{m}\,\rmx_{-\dot{-}} + p\bar{p}\,\rmx_{+\dot{+}}\big)^{\tilde{\jmath}} \\
&=\sum_{s=0}^{\tilde{\jmath}}\sum_{r=0}^{\min(s,\tilde{\jmath}-s)} \binom{\tilde{\jmath}}{2r}\binom{\tilde{\jmath}-2r}{s-r}\binom{2r}{r}\,(x^-)^{s-r}(x^+)^{\tilde{\jmath}-r-s}(x_\perp^2)^r\,(p\bar{m})^s (m\bar{p})^{\tilde{\jmath}-s}\\
&\quad+\sum_{s=0}^{\tilde{\jmath}-1}\sum_{t=0}^{\min(s,\tilde{\jmath}-s-1)} \binom{\tilde{\jmath}}{2t+1}\binom{\tilde{\jmath}-2t-1}{s-t}\binom{2t+1}{t+1}(x^-)^{s-t}(x^+)^{\tilde{\jmath}-t-s-1}(x_\perp^2)^t\times\\&\phantom{+\sum_{s=0}\sum_{\min(s,\tilde{\jmath}-s-1)}}\times
(m\bar{m}\,\rmx_{-\dot{-}}+p\bar{p}\,\rmx_{+\dot{+}})\,(p\bar{m})^s (m\bar{p})^{\tilde{\jmath}-s-1}
\,.
}[eq:expansionCharge2]
As before, both sums in $r$ and $t$ can be extended to any range. After
taking care of the remaining polarizations and performing the Fourier
transform with \eqref{eq:FTformulas} we again end up with sums in the form
of \eqref{eq:Sigmasum}. The result will be a $2\times 2$ matrix whose
entries are ratios of $\Gamma$ functions, which can be reduced to rational
functions of $\Delta,j$ and $s$. For the extreme cases $s=0$ and $s=j+1$
one needs to retain only the appropriate entry of this
matrix---respectively the upper left and the lower right---and discard the other ones. As an example we show the part of the formula that multiplies the coefficient $\cQObTQbO2$:
\eqn{
\anecE{\Delta}{j,1}{s}\Big|_{\cQObTQbO2} = -\frac{3\pi\lsp(-i)^{j+1}\lsp(\delta+1)(\delta+j+2)}{8\lsp(\delta+j-s+1)_3}
\left(
\begin{array}{cc}
\frac{\delta+j-s+3}{\delta+j-s+1} & \sqrt{\frac{s(j-s+1)}{(s+\delta)(\delta + j-s+1)}} \phantom{\Big|}\\
\sqrt{\frac{s(j-s+1)}{(s+\delta)(\delta + j-s+1)}} \phantom{\Big|} & \frac{(\delta+s-1)(\delta+j-s+1)}{(\delta+s)(\delta+j-s+4)}
\end{array}
\right)\,,
}[]
where now $\delta = \Delta -j/2 - 5/2$ with $\Delta$ the dimension of the
operator of spin $(\frac12 j,\frac12)$.

Clearly the same logic can be applied to more general cases $(\frac12
j,\frac12\jb)$ with $\jb$ fixed and $j$ arbitrary. It suffices to expand like in \eqref{eq:expansionCharge2} keeping terms with charge up to $\pm u$ where $u$ is the total number of invariants $\invI31$, $\invJ123$ and $\invJ231$ in the tensor structure under consideration. Then all steps follow in the same way, except that one may get sums more complicated than $\Sigma_{m,n}$.

\newsec{Tables}[appTabs]
\subsection{Ward identities}
\label{appendix:Ward}
\appendlist{wi}
\clearpage
\subsection{Expansion in components}
\label{appendix:expansion-tables}
\appendlist{nonsusy}
\clearpage

\end{appendices}

\bibliography{refs}

\end{document}